\documentclass[superscriptaddress,reprint, PRL]{revtex4-1}
\usepackage[english]{babel}
\usepackage{graphicx}
\usepackage[bbgreekl]{mathbbol}
\usepackage{amsfonts}
\usepackage{amssymb}
\usepackage{amsmath}
\usepackage{hyperref}
\usepackage{cancel}
\usepackage{colortbl}
\usepackage{siunitx}
\usepackage{soul}				
\usepackage{picture}
\usepackage{rotating}
\usepackage{textcomp}
\usepackage{textgreek} 			
\usepackage{transparent}

\usepackage{graphicx}
\usepackage{epstopdf}
\usepackage[abs]{overpic}
\usepackage{xcolor,varwidth}
\DeclareSymbolFontAlphabet{\mathbb}{AMSb}
\DeclareSymbolFontAlphabet{\mathbbl}{bbold}

\newcommand{\kbt}{k_{B}T}
\newcommand{\rhos}{\rho_{\rm s}}
\newcommand{\lb}{\lambda_{B}}
\newcommand{\ld}{\lambda_{D}}

\newcommand{\nn}{\nonumber\\}

\newcommand{\blue}[1]{\textcolor{blue}{#1}}

\begin{document}

\author{Mathijs Janssen}\email{mjanssen@is.mpg.de}
\affiliation{Institute for Theoretical Physics, Center for Extreme Matter and Emergent Phenomena,  Utrecht University, Princetonplein 5, 3584 CC Utrecht, The Netherlands}
\author{Elian Griffioen}
\affiliation{Van 't Hoff Laboratory for Physical and Colloid Chemistry, Debye Institute for Nanomaterials Science, Utrecht University, Padualaan 8, 3584 CH Utrecht, The Netherlands}
\author{P.~M. Biesheuvel}
\affiliation{Wetsus, European Centre of Excellence for Sustainable Water Technology, Oostergoweg 9, 8911 MA Leeuwarden, The Netherlands}
\author{Ren\'{e} van Roij}
\affiliation{Institute for Theoretical Physics, Center for Extreme Matter and Emergent Phenomena,  Utrecht University, Princetonplein 5, 3584 CC Utrecht, The Netherlands}
\author{Ben Ern\'{e}}\email{B.H.Erne@uu.nl}
\affiliation{Van 't Hoff Laboratory for Physical and Colloid Chemistry, Debye Institute for Nanomaterials Science, Utrecht University, Padualaan 8, 3584 CH Utrecht, The Netherlands}

\date{\today}

\begin{abstract}
Coulometric measurements on salt-water-immersed nanoporous carbon electrodes reveal, at a fixed voltage, a charge decrease with increasing temperature. During far-out-of-equilibrium charging of these electrodes, calorimetry indicates the production of both irreversible Joule heat and reversible heat, the latter being associated with entropy changes during electric double layer (EDL) formation in the nanopores. These measurements grant experimental access ---for the first time--- to the entropic contribution of the grand potential; for our electrodes, this amounts to roughly $25\%$ of the total grand potential energy cost of EDL formation at large applied potentials, in contrast with point-charge model calculations that predict $100\%$.
The coulometric and calorimetric experiments show a consistent picture of the role of heat and temperature in EDL formation
and provide hitherto unused information to test against EDL models.
\end{abstract}

\title{Coulometry and Calorimetry of Electric Double Layer Formation in Porous Electrodes}
\maketitle

Where surfaces of charged electrodes meet fluids that contain mobile ions, so-called electric double layers (EDLs) form that screen the electric surface charge by a diffuse cloud of counterionic charge.  
This EDL has been intensively studied for over a century and is of paramount importance to many processes in physical chemistry and soft matter physics. 
With the ongoing development of nanomaterials, nowadays electrodes can be made from porous carbon with internal surface areas exceeding 1000~m$^{2}$~g$^{-1}$. These porous electrodes can be immersed in a variety of electrolyte solutions or ionic liquids. A so-called electric double layer capacitor (EDLC) is then formed, whose high capacitance makes it a prime candidate for capacitive energy storage, energy conversion \cite{brogioli2009extracting, hartel2015heat}, and water desalination \cite{hemmatifar2015two, kim2015enhanced, rubin2016induced}. 
In these porous electrodes, solvated ions have a size similar to that of their confining geometry; hence, a realistic theory must at least address both the electrostatics and the packing of the ions. Simulations and \textit{in situ} analytical techniques \cite{wang2013situ} have revealed a wealth of phenomena in EDLCs \cite{kornyshev2014three}, including overscreening \cite{Bazant:2011aa}, ion desolvation  \cite{merlet2012molecular, bankura2013hydration}, 
in-plane structural transitions \cite{merlet2014electric}, layered packings of counterionic charge at high surface potentials \cite{hartel2015fundamental}, and, relatedly,  oscillations in the EDL capacitance with decreasing pore width \cite{feng2011supercapacitor,wu2011complex,jiang2011oscillation}. 
Unfortunately, the gap between (computationally demanding) first-principles models and experimental measurements on the charging behavior of porous electrodes is far from closed, with many questions remaining regarding the precise screening mechanisms at play \cite{smith2016electrostatic}. While our understanding of the EDL is based mainly on {\it isothermal} numerical and experimental methods, recent work has revealed an interplay between temperature, heat, and entropy in the EDL. In particular, both model calculations \cite{Janssen:2014aa} and experiments \cite{hartel2015heat} indicate that the surface potential of an electrode with a fixed high surface charge should rise by about $1$ mV K$^{-1}$ with increasing temperature. 
Conversely, EDL formation under \textit{adiabatic} settings induces a thermal response that has been largely overlooked \cite{janssen2017reversible}.
During charging, ions and solvent molecules order into the EDL; hence, the configurational contribution to the total phase-space volume  decreases. During \textit{isentropic} charging, this decrease must be balanced by an equal and opposite increase in the momentum contribution: the electrolyte solution heats up.
As the source of reversible temperature variations is nonzero only within the EDL, temperature variations scale inversely with the average electrode pore size. The development of  high-surface-to-volume-ratio electrodes (for, e.g., supercapacitors) was therefore required before the small temperature variations were detected experimentally \cite{schiffer2006heat}.

The temperature-dependent phenomena mentioned above are (in principle) measurable and therefore open new possibilities for experiments against which EDL theories can be tested. In this Letter on the temperature-dependent EDL we present data of two experiments involving water-immersed porous carbon electrodes, which is an important system for water desalination and blue-energy devices \cite{brogioli2009extracting}. 
Conceptually, the two experiments are each other's ``opposites": while the first experiment 
involves a temperature-induced charge variation, the second experiment studies a charge-induced temperature variation. 
More specifically, with a potentiostatic  \textit{coulometry} experiment we determine the temperature dependence of the equilibrium charge of our blocking electrodes at fixed potential. 
Two thermodynamic identities then allow us to predict not only the heat required to flow out of the cell upon isothermal charging, but also the thermal response to adiabatic charging. 
In the second,  \textit{calorimetric} experiment, we probe the thermal response of the capacitor to far-from-equilibrium charging.  
We then distinguish irreversible Joule heat from reversible heat, which we identify with the entropic contribution to the grand potential. While previous work built theoretical models for the latter quantity \cite{overbeek1990role, kralj1996simple}, to our knowledge, this Letter presents its first experimental determination. 
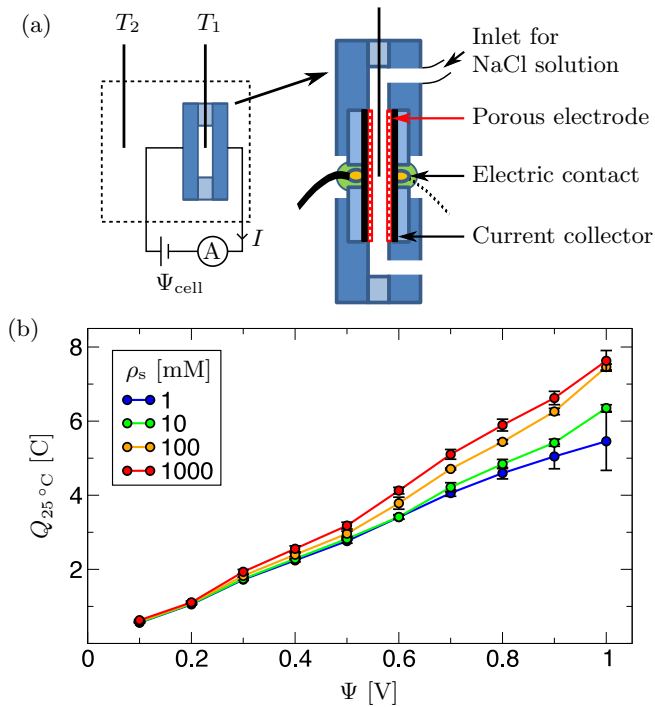
\begin{figure}
\def\svgwidth{0.27\textwidth}{ 
 \providecommand\color[2][]{%
    \errmessage{(Inkscape) Color is used for the text in Inkscape, but the package 'color.sty' is not loaded}%
    \renewcommand\color[2][]{}%
  }%
  \providecommand\transparent[1]{%
    \errmessage{(Inkscape) Transparency is used (non-zero) for the text in Inkscape, but the package 'transparent.sty' is not loaded}%
    \renewcommand\transparent[1]{}%
  }%
  \providecommand\rotatebox[2]{#2}%
  \ifx\svgwidth\undefined%
    \setlength{\unitlength}{196.86688849bp}%
    \ifx\svgscale\undefined%
      \relax%
    \else%
      \setlength{\unitlength}{\unitlength * \real{\svgscale}}%
    \fi%
  \else%
    \setlength{\unitlength}{\svgwidth}%
  \fi%
  \global\let\svgwidth\undefined%
  \global\let\svgscale\undefined%
  \makeatother%
  \begin{picture}(1,0.82)%
    \put(-0.15,0){\includegraphics[width=\unitlength]{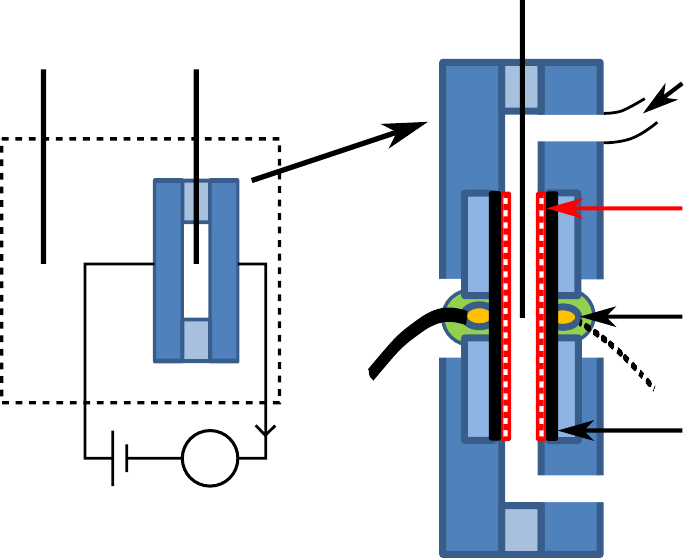}}%
    \put(0.13,0.1254167){\color[rgb]{0,0,0}\makebox(0,0)[lb]{\smash{A}}}%
    \put(-0.11,0.75528361){\color[rgb]{0,0,0}\makebox(0,0)[lb]{\smash{$T_{2}$}}}%
    \put(0.115,0.75528361){\color[rgb]{0,0,0}\makebox(0,0)[lb]{\smash{$T_{1}$}}}%
    \put(0.0,0.04){\color[rgb]{0,0,0}\makebox(0,0)[lb]{\smash{$\Psi_{\rm cell}$}}}%
    \put(0.26390354,0.16003556){\color[rgb]{0,0,0}\makebox(0,0)[lb]{\smash{$I$}}}%
    \put(0.87,0.715){Inlet for }
    \put(0.87,0.64){NaCl solution}
    \put(0.87,0.495){Porous electrode}
    \put(0.87,0.335){Electric contact}
    \put(0.87,0.17){Current collector}
    \put(-0.37,0.75){(a)}
  \end{picture}
  }
\def\svgwidth{0.43\textwidth}{
  \providecommand\color[2][]{%
    \errmessage{(Inkscape) Color is used for the text in Inkscape, but the package 'color.sty' is not loaded}%
    \renewcommand\color[2][]{}%
  }%
  \providecommand\transparent[1]{%
    \errmessage{(Inkscape) Transparency is used (non-zero) for the text in Inkscape, but the package 'transparent.sty' is not loaded}%
    \renewcommand\transparent[1]{}%
  }%
  \providecommand\rotatebox[2]{#2}%
  \ifx\svgwidth\undefined%
    \setlength{\unitlength}{400.57144165bp}%
    \ifx\svgscale\undefined%
      \relax%
    \else%
      \setlength{\unitlength}{\unitlength * \real{\svgscale}}%
    \fi%
  \else%
    \setlength{\unitlength}{\svgwidth}%
  \fi%
  \global\let\svgwidth\undefined%
  \global\let\svgscale\undefined%
  \makeatother%
 \begin{picture}(1,0.68)%
    \put(0.025,0.025){\includegraphics[width=\unitlength]{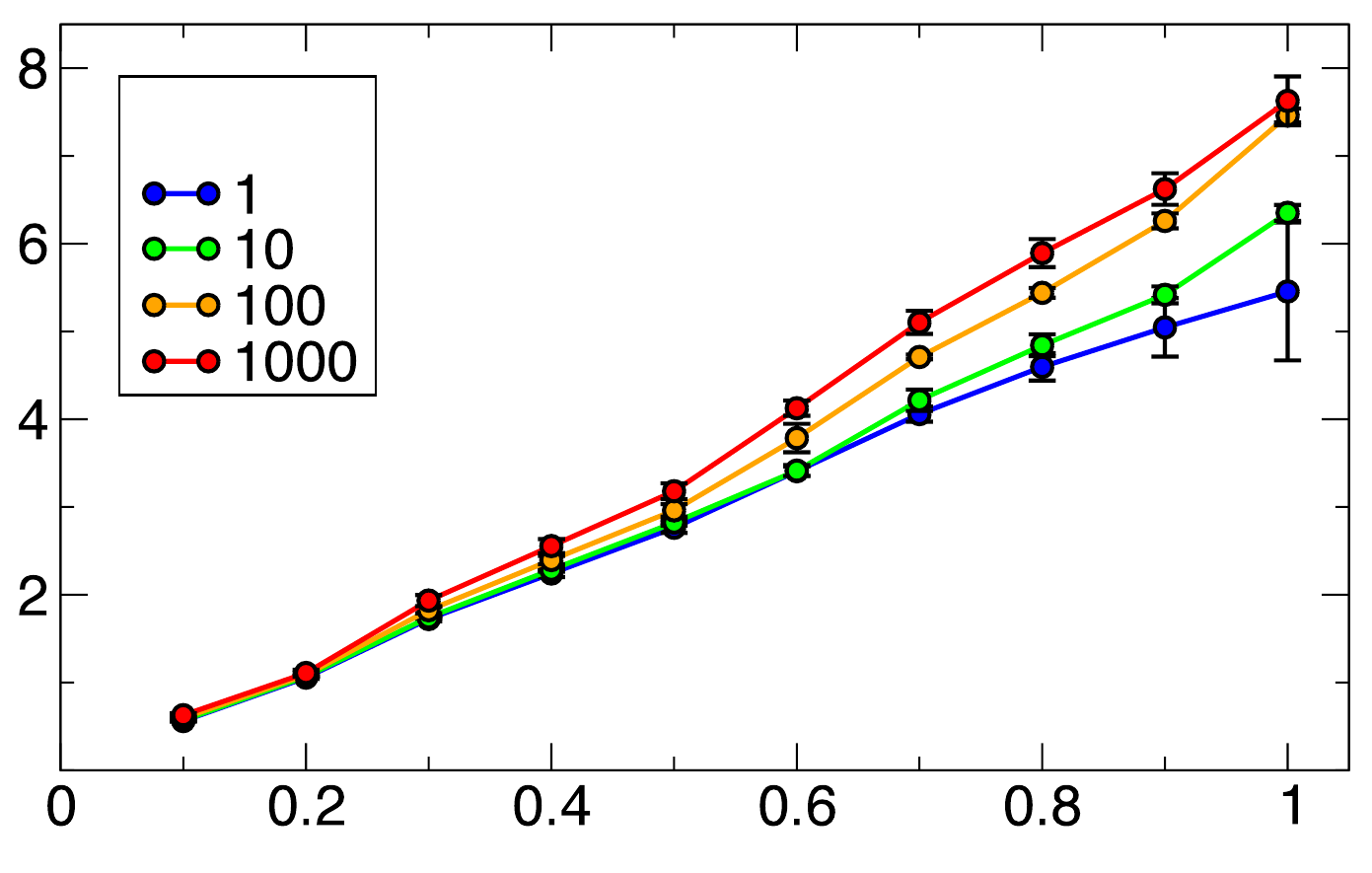}}%
    \put(0.135,0.56){\color[rgb]{0,0,0}\makebox(0,0)[lb]{\smash{$\rhos$ [mM]}}}%
    \put(0.00,0.28){\color[rgb]{0,0,0}\makebox(0,0)[lb]{\smash{\begin{rotate}{90} $Q_{\SI{25}{\degreeCelsius}}$ [C]\end{rotate}}}}%
    \put(0.50356632,0.00364713){\color[rgb]{0,0,0}\makebox(0,0)[lb]{\smash{$\Psi$ [V]}}}%
    \put(-0.07,0.63){(b)}
  \end{picture}%
}
\caption{\label{fig1}%
(a) The experimental setup to measure temperature and charge during (dis)charging of porous carbon electrodes (0.18 g each) in NaCl solution. 
(b)  The dependence of the total charge $Q$ on $\Psi$ is shown at $\rhos=1, 10, 100, 1000$ mM. Error bars are based on two or more duplicate measurements.}
\end{figure}

Experiments were performed using a homebuilt electrochemical cell submerged in a thermostatic water bath maintained at $\SI{25}{\degreeCelsius}$ (Julabo F25). 
The cell, depicted in Fig.~\ref{fig1}(a), had in- and outlets for a degassed aqueous NaCl solution at salt concentration $\rhos$ and a Pt100 temperature probe (0.5 mm diameter, miniature RTD sensor from TC Direct) that measured the temperature  $T_{1}$ within the cell. The tip of the probe was centered halfway between two concentric and parallel electrodes, separated from each other by \mbox{2.2 mm}.  The electrodes were disks of 25 mm in diameter cut from sheets of porous carbon: 0.5 mm thick, with a density of 0.58~g~mL$^{-1}$, a porosity of $65\%$, weighing $0.18$~g each, and a  Brunauer-Emmett-Teller-area of 1330~m$^{2}$~g$^{-1}$ from nitrogen adsorption \cite{kim2015enhanced}. These disks were glued with conductive silver epoxy onto graphite current collector disks. In turn, the electrodes and current collectors were mounted in a glass casing, with holes allowing copper wires to make electric contact with both collectors using silver epoxy glue. There, nonconducting glue applied onto the copper and silver epoxy prevented direct contact with the thermostatic water bath. Cell voltages were applied and electrical currents were measured using a potentiostat-galvanostat PGSTAT100 from Metrohm Autolab. The temperature difference $\Delta T=T_{1}-T_{2}$  was measured compared to a second Pt100 temperature probe ($T_{2}$) immersed directly in the thermostatic bath.  Without an applied potential, $\Delta T$ was constant within $\SI{0.007}{\degreeCelsius}$ for several days.

In both experiments, at time $t=t_{0}$ the cell voltage $\Psi_{\mathrm{cell}}$ was changed stepwise from 0 V to different maximum values $\Psi$ not exceeding 1~V. 
The total charge $Q_{T_{1}}(\Psi)=\int_{t_{0}}^{t_{1}}Idt$ on the electrodes' surface at temperature $T_{1}$, shown for several salt concentrations in Fig.~\ref{fig1}(b), was found by integrating the current $I$ until $t_{1}$, 4 hours later. 

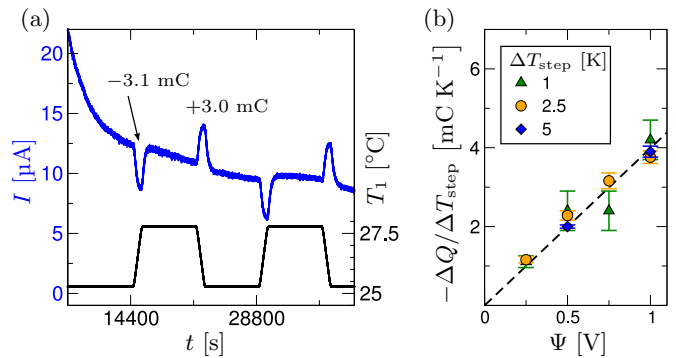
\begin{figure}[b]
\centering
\def\svgwidth{0.49\textwidth}{
\providecommand\color[2][]{%
    \errmessage{(Inkscape) Color is used for the text in Inkscape, but the package 'color.sty' is not loaded}%
    \renewcommand\color[2][]{}%
  }%
  \providecommand\transparent[1]{%
    \errmessage{(Inkscape) Transparency is used (non-zero) for the text in Inkscape, but the package 'transparent.sty' is not loaded}%
    \renewcommand\transparent[1]{}%
  }%
  \providecommand\rotatebox[2]{#2}%
  \ifx\svgwidth\undefined%
    \setlength{\unitlength}{623.77611932bp}%
    \ifx\svgscale\undefined%
      \relax%
    \else%
      \setlength{\unitlength}{\unitlength * \real{\svgscale}}%
    \fi%
  \else%
    \setlength{\unitlength}{\svgwidth}%
  \fi%
  \global\let\svgwidth\undefined%
  \global\let\svgscale\undefined%
  \makeatother%
  \begin{picture}(1,0.5)%
    \put(-0.02,0){\includegraphics[width=\unitlength]{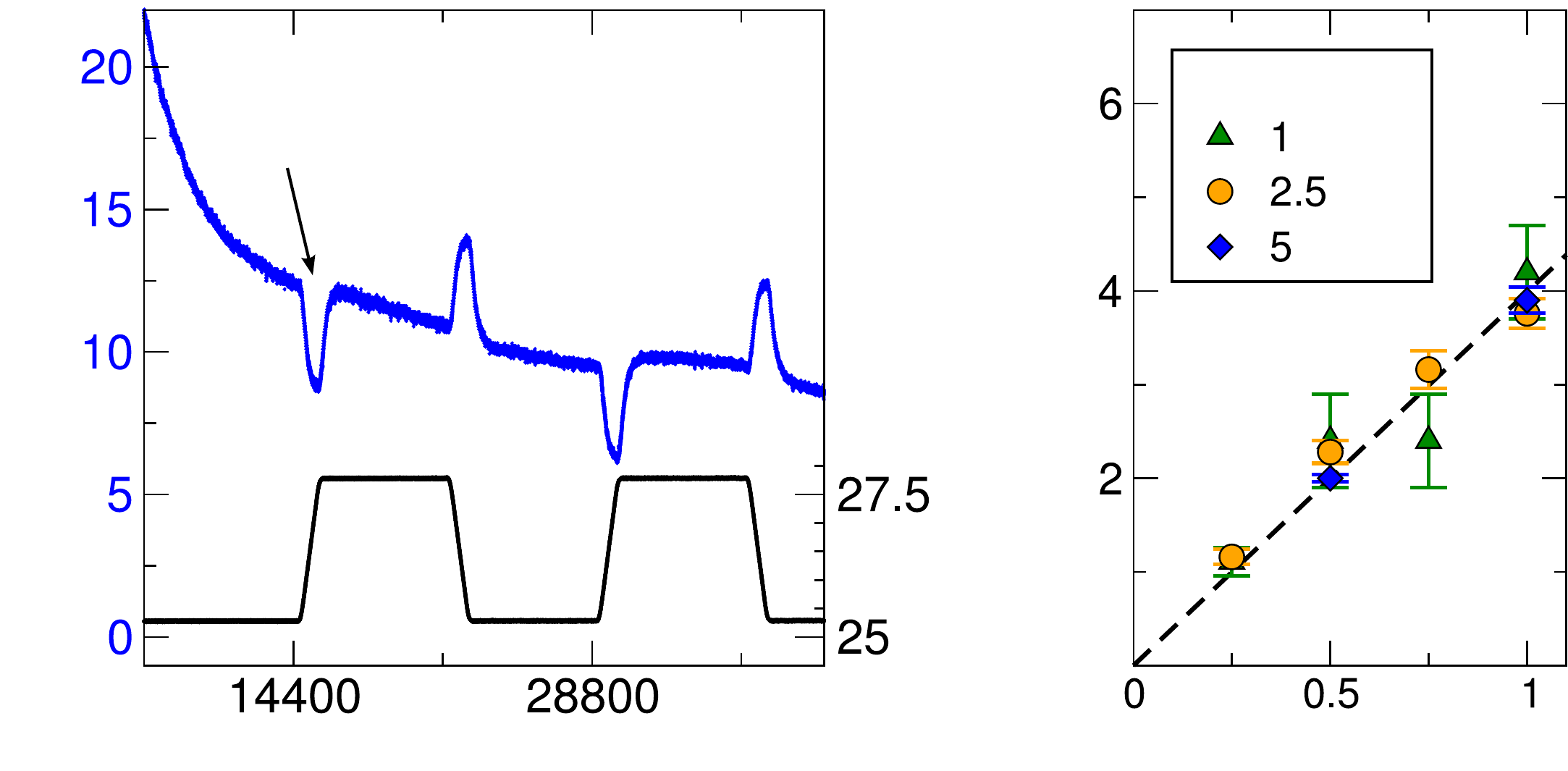}}%
   \put(0.0,0.49){\color[rgb]{0,0,0}\makebox(0,0)[lb]{\smash{(a)}}}%
    \put(0.61,0.49){\color[rgb]{0,0,0}\makebox(0,0)[lb]{\smash{(b)}}}%
    \put(0.25,0.00){\color[rgb]{0,0,0}\makebox(0,0)[lb]{\smash{$\displaystyle t$ [s]}}}%
    \put(0.8,0.0){\color[rgb]{0,0,0}\makebox(0,0)[lb]{\smash{$\Psi$ [V]}}}%
    \put(0.655,0.07){\color[rgb]{0,0,0}\makebox(0,0)[lb]{\smash{\begin{rotate}{90} $-\Delta Q/\Delta T_{\rm step}$ \big[mC~K$^{-1}$\big]\end{rotate}}}}%
    \put(0.015,0.22){\color[rgb]{0,0,0}\makebox(0,0)[lb]{\smash{\begin{rotate}{90} \small \blue{$ I$ [\textmu A]}\end{rotate}}}}%
    \put(0.55,0.22){\color[rgb]{0,0,0}\makebox(0,0)[lb]{\smash{\begin{rotate}{90} \small$T_{1}~[\SI{}{\degreeCelsius}]$\end{rotate}}}}%
      \put(0.25,0.36){\color[rgb]{0,0,0}\makebox(0,0)[lb]{\smash{\scriptsize $+3.0$~mC}}}%
    \put(0.13051436,0.4){\color[rgb]{0,0,0}\makebox(0,0)[lb]{\smash{\scriptsize $-3.1$~mC}}}%
    \put(0.74,0.43){\color[rgb]{0,0,0}\makebox(0,0)[lb]{\smash{\scriptsize$\Delta T_{\rm step}$ [K]}}}%
  \end{picture}%
}
\caption{\label{fig2}%
(a) In the coulometric experiment, the current (blue)  initially exhibits a monotonic decay, onto which positive and negative peaks are superimposed when the electrolyte temperature (black) is decreased and increased.  Plotted are data for $\Delta T_{\rm step}=\SI{2.5}{\degreeCelsius}$, $\rhos=1$~M, and $\Psi=0.25$~V. 
(b) The temperature-scaled equilibrium charge difference $\Delta Q/\Delta T_{\rm step}$ is shown at different potential and temperature steps.} 
\end{figure}

In the coulometric experiment [see Fig.~\ref{fig2}(a)], after the voltage rise from $0$~V to $\Psi$ at $t_{0}$ and approximately 4 hours of subsequent equilibration (all at $25.3\pm\SI{0.1}{\degreeCelsius}$), the temperature was alternated at fixed voltage $\Psi$ between two temperatures differing by $\Delta T_{\rm step}$, with ample time for equilibration in between. Here, the temperature changes ---with a temporal spacing of 2 hours--- are accomplished at a thermal sweeping rate of $\SI{10}{\degreeCelsius}$ per hour.
By time integrating peaks in the current $I$, we determined the excess charge $\Delta Q$ that flowed because of the temperature-induced change in the capacitance. For instance, at $\Psi=1$ V and $\Delta  T_{\rm step}=\SI{2.5}{\degreeCelsius}$, the charge decreased from $Q_{\SI{25.3}{\degreeCelsius}}=8.11$~C by an amount of $\Delta Q=Q_{\SI{27.8}{\degreeCelsius}}-Q_{\SI{25.3}{\degreeCelsius}}=-9.40$~mC, i.e., a relative decrease of about $0.05\%$~K$^{-1}$. 
The charge decrease scaled to the temperature step is shown in Fig.~\ref{fig2}(b) for several  $\Delta T_{\rm step}$ and $\Psi$. We see that $\Delta Q$ scales linearly with both $\Delta T_{\rm step}$ and $\Psi$. We write $\left(\partial \Delta Q/\partial T\right)_{\Psi}=\alpha\Psi$, and determine $\alpha=-4.02\pm0.07$~mC~V$^{-1}$~K$^{-1}$ from a linear fit to the data in Fig.~\ref{fig2}(b).
With $\alpha$ at hand, we can predict the heat $\mathbb{Q}$ required to flow into the device upon  isothermal charging, and also the adiabatic temperature rise $\Delta T_{\rm adiab}$ upon charging the cell, \textit{had it been thermally insulated}.
The required thermodynamic identities for $\mathbb{Q}$ and $\Delta T_{\rm adiab}$ were previously derived in Refs.~\cite{hartel2015heat} and \cite{Janssen:2014aa}, respectively, and take simplified forms for the capacitor of interest (see Appendices~\ref{heatflow} and \ref{sec_adiabatictemperaturerise}, respectively). 
In particular, the isothermal heat amounts to $\mathbb{Q}=\alpha T\Psi^2/2$; hence, \mbox{$\mathbb{Q}=-0.60 \pm 0.01 $}~J~at $\Psi=$~1 V; i.e., a positive amount of heat flows \textit{out} of the capacitor during isothermal charging, while an equal amount of heat should flow \textit{into} the capacitor during the mirror discharging process. %
In the presence of insulating walls, heat cannot exit the cell and the capacitor would exhibit a temperature rise of $\Delta T_{\rm adiab}=\alpha T\Psi^2/(2\mathbb{C}_{p})$ upon charging ---with $\mathbb{C}_{p}=10.9\pm0.1$  J~K$^{-1}$ being the heat capacity of the cell determined in Appendix~\ref{sec:alternativecalorimetry}--- and an equal-sized cooling effect for the mirror discharging process. At $\Psi=$~1 V, this amounts to $\Delta T_{\rm adiab}=0.055\pm0.002$~K. 

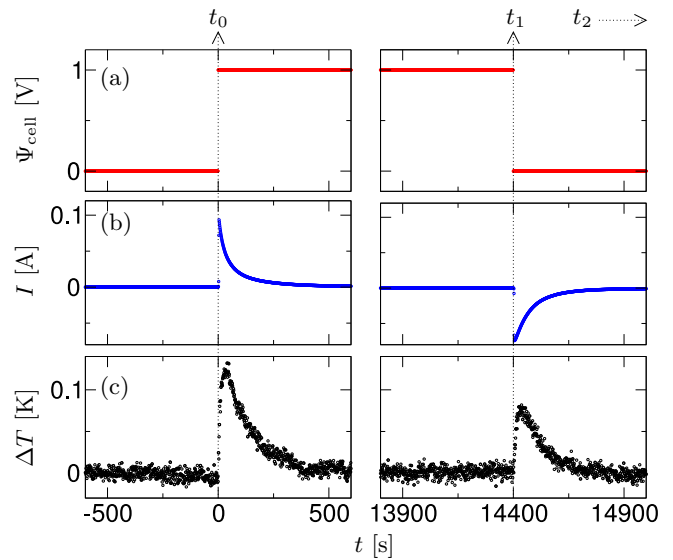
\begin{figure}
\centering
\def\svgwidth{0.45\textwidth}{ 
  \providecommand\color[2][]{%
    \errmessage{(Inkscape) Color is used for the text in Inkscape, but the package 'color.sty' is not loaded}%
    \renewcommand\color[2][]{}%
  }%
  \providecommand\transparent[1]{%
    \errmessage{(Inkscape) Transparency is used (non-zero) for the text in Inkscape, but the package 'transparent.sty' is not loaded}%
    \renewcommand\transparent[1]{}%
  }%
  \providecommand\rotatebox[2]{#2}%
  \ifx\svgwidth\undefined%
    \setlength{\unitlength}{505.6735bp}%
    \ifx\svgscale\undefined%
      \relax%
    \else%
      \setlength{\unitlength}{\unitlength * \real{\svgscale}}%
    \fi%
  \else%
    \setlength{\unitlength}{\svgwidth}%
  \fi%
  \global\let\svgwidth\undefined%
  \global\let\svgscale\undefined%
  \makeatother%
  \begin{picture}(1,0.91)%
    \put(0.02,0){\includegraphics[width=\unitlength]{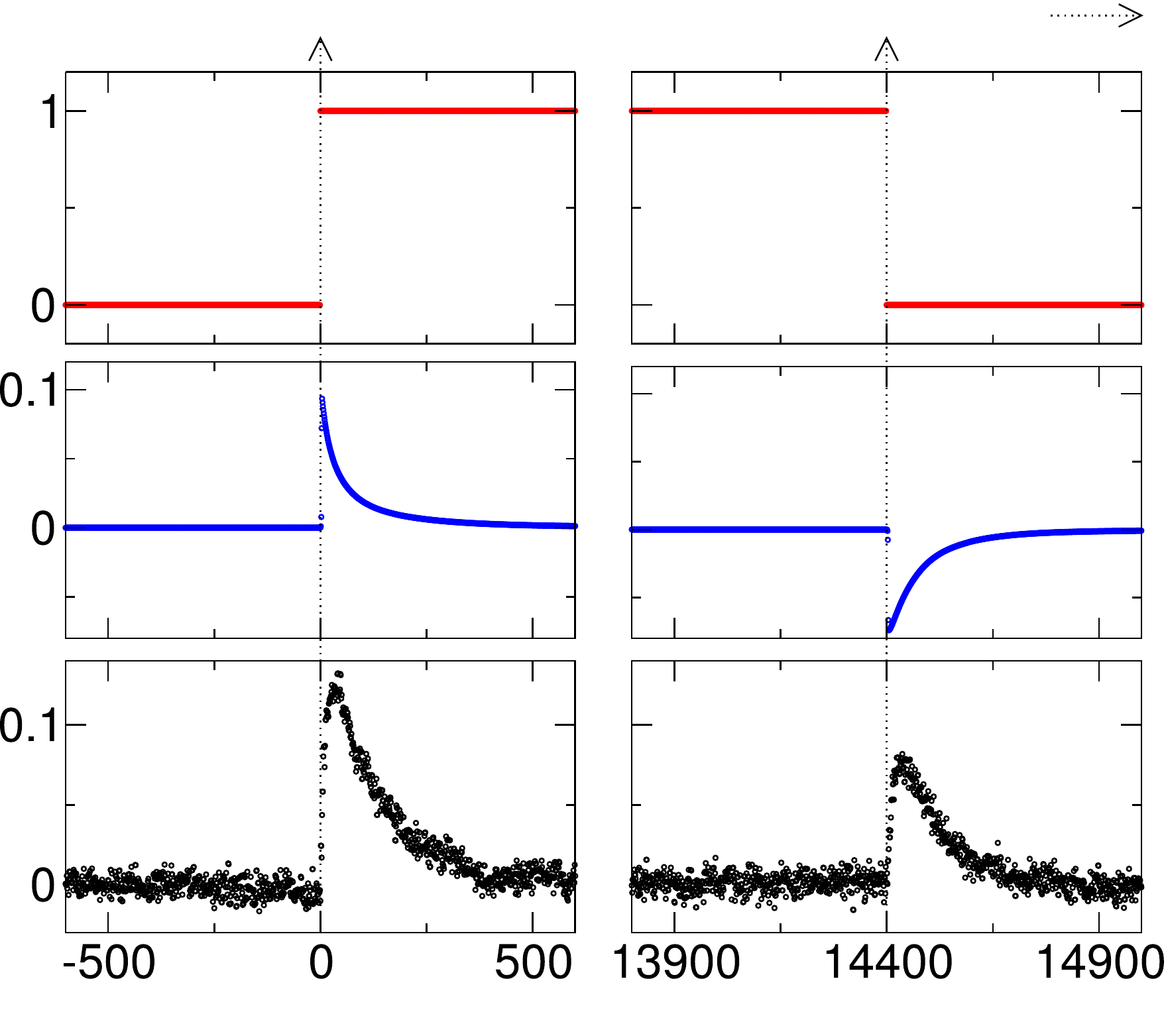}}%
    \put(0.1,0.77){\color[rgb]{0,0,0}\makebox(0,0)[lb]{\smash{(a)}}}%
    \put(0.1,0.53){\color[rgb]{0,0,0}\makebox(0,0)[lb]{\smash{(b)}}}%
    \put(0.1,0.26){\color[rgb]{0,0,0}\makebox(0,0)[lb]{\smash{(c)}}}%
    \put(0.28,0.87){\color[rgb]{0,0,0}\makebox(0,0)[lb]{\smash{$\displaystyle t_{0}$}}}%
    \put(0.77,0.87){\color[rgb]{0,0,0}\makebox(0,0)[lb]{\smash{$\displaystyle t_{1}$}}}%
    \put(0.88,0.87){\color[rgb]{0,0,0}\makebox(0,0)[lb]{\smash{$\displaystyle t_{2}$}}}%
    \put(-0.01,0.64){\color[rgb]{0,0,0}\makebox(0,0)[lb]{\smash{\begin{rotate}{90} $\displaystyle \Psi_{\mathrm{cell}}$ [V]\end{rotate}}}}%
    \put(-0.01,0.42){\color[rgb]{0,0,0}\makebox(0,0)[lb]{\smash{\begin{rotate}{90} $\displaystyle I$ [A]\end{rotate}}}}%
   \put(-0.01,0.14){\color[rgb]{0,0,0}\makebox(0,0)[lb]{\smash{\begin{rotate}{90} $\displaystyle \Delta T$ [K]\end{rotate}}}}%
    \put(0.52,0.0){\color[rgb]{0,0,0}\makebox(0,0)[lb]{\smash{$\displaystyle t$ [s]}}}%
  \end{picture}%
}
\caption{\label{fig3}%
In the calorimetric experiment, the cell voltage $\Psi_{\rm cell}$ (a) was switched to $\Psi$ at $t_0$ and 0 V was applied at $t_1$. The current $I$ (b) and temperature difference $\Delta T$ (c) were measured simultaneously. Plotted are data for $\Psi=1$ V and $\rhos=1$~M.}
\end{figure}

In the second, calorimetric, experiment, after the initial voltage rise and subsequent equilibration, at $t=t_{1}$ we switch the voltage [see Fig.~\ref{fig3}(a)] back to 0 and, while measuring $\Delta T$, wait another 4 hours until $t=t_{2}$ when the current $I$  [Fig.~\ref{fig3}(b)] has essentially dropped to 0 (the relaxation time scales are discussed in Appendix~\ref{sec:app_timescales}). Since this (dis)charging process is far from equilibrium, the $\Delta T$ peaks in Fig.~\ref{fig3}(c) are caused both by reversible rearrangements within the EDL and by irreversible strictly positive Joule heating. 
Clearly, the exothermic Joule heat dominates in both directions of charge transfer. Nevertheless, heat is more exothermic upon charging ($t_{0}\to t_{1}$)  than upon discharging ($t_{1}\to t_{2}$); the difference is ascribed to the reversible part of the heat exchange, which is exothermic upon charging and endothermic upon discharging. 

At any moment in this charging and discharging cycle, the measured temperature difference $\Delta T$ could, in principle, be found from the heat equation \cite{de1951thermodynamics, biesheuvel2014negative},
\begin{equation}\label{eq:microheatequation}
\varrho c_{p}\partial_{t} T= \kappa \nabla^2 T+{\bf I}\cdot{\bf E}
\end{equation}  
describing the time evolution of the temperature $T({\bf x}, t)$ at any point ${\bf x}$ in the thermostatic bath $\mathcal{V}_{\rm bath}$, the solid $\mathcal{V}_{s}$ (carbon electrodes, glass casing, circuitry, etc.), and the electrolyte solution $\mathcal{V}_{\rm el}$ regions of the cell. Here, the ionic current ${\bf I}$ and electric field ${\bf E}$ are spatially varying and (possibly) nonzero only within $\mathcal{V}_{\rm el}$. Moreover, all material properties appearing in Eq.~\eqref{eq:microheatequation} are locally defined; the specific heat capacity $c_{p}$, mass density $\varrho$, and heat conductivity $\kappa$ take different values in the different parts of the system ($\mathcal{V}_{\rm bath}, \mathcal{V}_{s}, \mathcal{V}_{\rm el}$). 
Solving Eq~\eqref{eq:microheatequation} for the complete three-dimensional geometry of the cell (including the porous network) and bath is out of the scope of our study and a simplified analysis of the experiments is performed. 

In our analysis, the temperature is assumed to be homogeneous within the complete cell, with a steep drop of $\Delta T$ between the cell and the thermostatic bath. The temperature difference $\Delta T$ between the cell and the thermostatic bath changes when heat is added at another rate than heat is lost to the environment. 
If the heat flow to the environment is assumed to be linear with $\Delta T$, as in Newton's law of cooling, the temperature difference $\Delta T$ is governed by
\begin{equation}\label{eq:heatequation}
\mathbb{C}_{p}\frac{d\Delta T}{dt}=-K\Delta T+\dot{\Pi}_{\rm tot},
\end{equation} 
where we assume $d\Delta T/dt = dT_{1}/dt$, thanks to a practically constant $T_{2}$, and with $\dot{\Pi}_{\rm tot}=\int_{\mathcal{V}_{\rm el}} d{\bf x}~{\bf I}\cdot{\bf E}$ being the total heating rate in J~s$^{-1}$, and $K$ being the heat transfer coefficient in J~s$^{-1}$~K$^{-1}$ of the cell-bath interface.

Equation~(\ref{eq:microheatequation}) stems from an internal energy balance; 
the source term ${\bf I}\cdot{\bf E}$ captures the exchange between electric field energy and the solution's thermal energy as the electric field performs work on the electrolyte solution. This source term can be split up in reversible ($\sim I$) and irreversible contributions ($\sim I^2$), respectively, with only the reversible term persisting in the limit of slow charging \cite{d2014first, janssen2017reversible}. 
Likewise, for the total heat production rate appearing in Eq.~(\ref{eq:heatequation}) we similarly write 
\begin{equation}\label{eq:productionsplitting}
\int_{t_{0}}^{t_{2}}\dot{\Pi}_{\rm tot}dt\equiv\Pi_{\rm tot}\equiv\Pi_{\rm rev}^{\mathrm{ch}}+\Pi_{\rm irr}^{\mathrm{ch}}+\Pi_{\rm rev}^{\mathrm{dis}}+\Pi_{\rm irr}^{\mathrm{dis}},
\end{equation} 
with reversible (rev) and irreversible (irr) heat contributions, during charging (ch, $t_{0}\le t< t_{1}$) and discharging (dis, $t_{1}\le t\le t_{2}$). The right-hand side of Eq.~(\ref{eq:productionsplitting}) simplifies since the reversible heat during charging opposes that during discharging by definition ($\Pi_{\rm rev}^{\mathrm{ch}}=-\Pi_{\rm rev}^{\mathrm{ch}}$).  
To appreciate the significance of the term $\Pi_{\rm rev}^{\mathrm{ch}}$ (and likewise, $\Pi_{\rm rev}^{\mathrm{dis}}$), we note that reversible work performed by the electric field on the electrolyte solution equals the entropic contribution $\Omega_{\rm ent}$ to the grand potential  $\Omega$, as was pointed out by Overbeek \cite{overbeek1990role}: ``during the charging process ... the solution part of the double layer arranges itself automatically, i.e., with zero contribution to the free energy, and thus the change in entropy and the electrical work done [on the solution, $\int_{\mathcal{V}_{\rm el}} d\textbf{x}\int_{0}^{\psi}\psi dq$, with $\psi(\textbf{x})$ and $q(\textbf{x})$ being the local electrostatic potential and charge density, respectively] just compensate one another."  Hence, $\Pi_{\mathrm{rev}}^{\mathrm{ch}}$ is to be identified with the entropic contribution $\Omega_{\mathrm{ent}}$ to the grand potential  $\Omega$ of the cell during EDL buildup, which, as far as we know, has not previously been measured experimentally. 
Interestingly,  the equality $\Pi_{\rm rev}^{\mathrm{ch}}=\Omega^{\mathrm{GC}}_{\mathrm{ent}}$ can be derived explicitly (see Appendix~\ref{heatproduction}) within the classical Gouy-Chapman (GC) EDL model, for which an analytical expression for $\Omega_{\mathrm{ent}}$ is known. 

To find $\Pi_{\mathrm{rev}}^{\mathrm{ch}}$, we first need to determine the empirical parameter $K$, which is done in Appendix \ref{sec:calibration} via two calibration methods, both employing known quantities of Joule heat. 
One of these methods used the geometry with carbon electrodes (c.e.) as described so far, and determined $K^{\textrm{c.e.}} = 0.239\pm 0.003$~J~s$^{-1}$~K$^{-1}$ directly from the data presented in Fig.~\ref{fig3}. 
In this setup, most of the Joule heat could not have been produced in the electrolyte solution, as the total resistance went up from 13 to merely~70 $\Omega$ upon going from 1 M to 1 mM NaCl, despite a factor of 1000 in electrolyte conductivity.
Instead, the Joule heat was probably produced at the electrical contacts between the porous carbon electrodes and the external circuit \cite{kim2015enhanced}. 
In the other calibration method, we replaced the electrodes with heating elements (h.e.) where heat is generated in a wire of known resistance. This gave $K^{\textrm{h.e.}} = 0.179\pm0.001$~J~s$^{-1}$~K$^{-1}$. Even though the latter cell setup has a (slightly) different heat conductance and conductivity, we deem this calibration method with heating elements superior because (1) Joule heat is determined more accurately and (2) the heat with which is calibrated is generated at the same location as the reversible heat, the quantity we set out to find. 

\begin{figure}
\def\svgwidth{0.48\textwidth}{
  \providecommand\color[2][]{%
    \errmessage{(Inkscape) Color is used for the text in Inkscape, but the package 'color.sty' is not loaded}%
    \renewcommand\color[2][]{}%
  }%
  \providecommand\transparent[1]{%
    \errmessage{(Inkscape) Transparency is used (non-zero) for the text in Inkscape, but the package 'transparent.sty' is not loaded}%
    \renewcommand\transparent[1]{}%
  }%
  \providecommand\rotatebox[2]{#2}%
  \ifx\svgwidth\undefined%
    \setlength{\unitlength}{408bp}%
    \ifx\svgscale\undefined%
      \relax%
    \else%
      \setlength{\unitlength}{\unitlength * \real{\svgscale}}%
    \fi%
  \else%
    \setlength{\unitlength}{\svgwidth}%
  \fi%
  \global\let\svgwidth\undefined%
  \global\let\svgscale\undefined%
  \makeatother%
  \begin{picture}(1,1.27)%
    \put(-0.01,-0.02){\includegraphics[width=\unitlength]{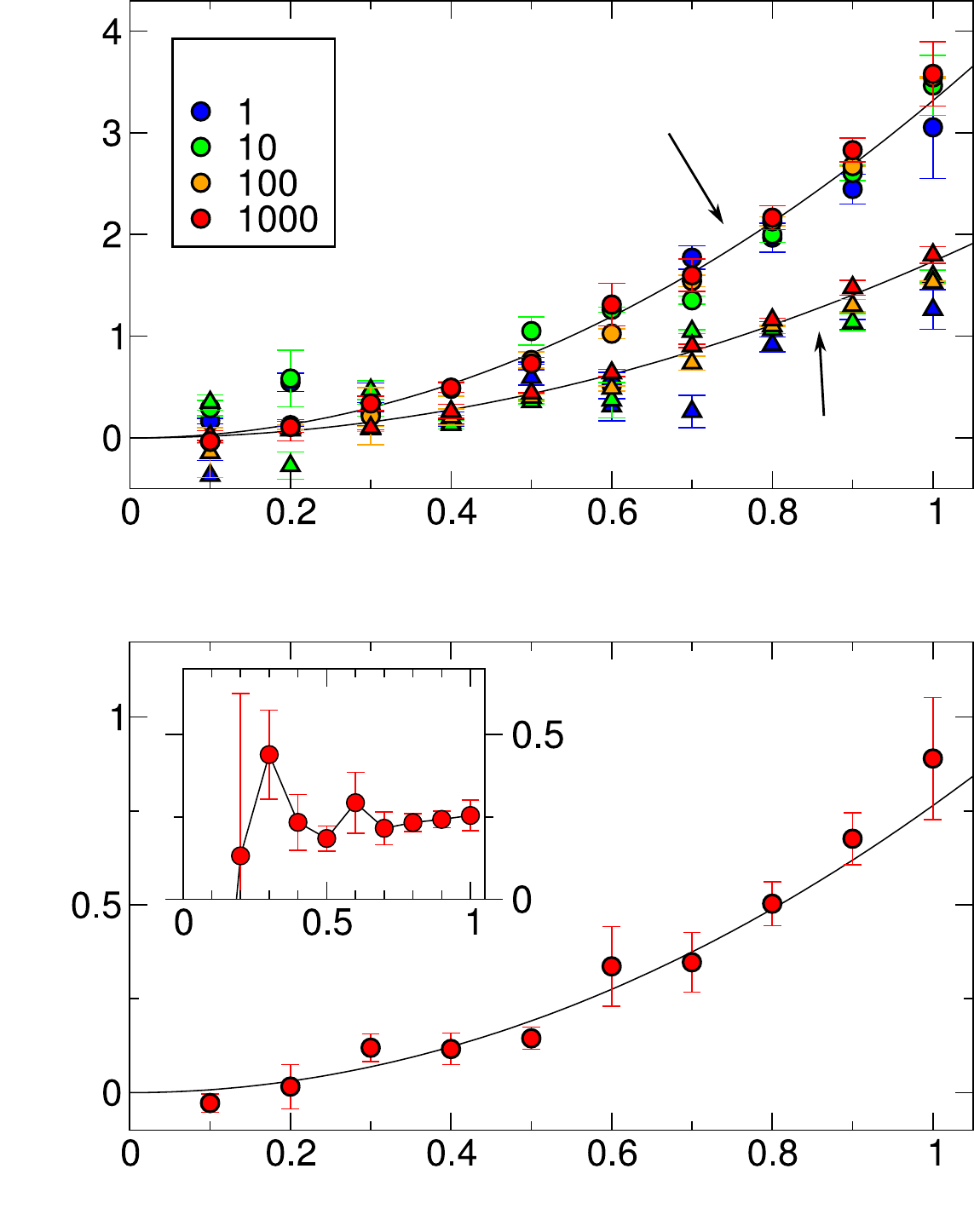}}%
    \put(0.0,1.23){(a)}
    \put(0.0,0.57){(b)}
    \put(0.19,1.165){\color[rgb]{0,0,0}\makebox(0,0)[lb]{\smash{$\rhos$ [mM]}}}%
    \put(0.035,0.93){\color[rgb]{0,0,0}\makebox(0,0)[lb]{\smash{\begin{rotate}{90}$\displaystyle  \Pi_{\mathrm{tot}}~\big[$J$\big]$\end{rotate}}}}%
    \put(0.49,0.65){\color[rgb]{0,0,0}\makebox(0,0)[lb]{\smash{$\Psi$~[V]}}}%
    \put(0.49,0.00){\color[rgb]{0,0,0}\makebox(0,0)[lb]{\smash{$\Psi$~[V]}}}%
    \put(0.27,0.24){\color[rgb]{0,0,0}\makebox(0,0)[lb]{\smash{$\Psi$~[V]}}}%
    \put(0.64,1.13){\color[rgb]{0,0,0}\makebox(0,0)[lb]{\smash{$\displaystyle  \Pi_{\mathrm{tot}}^{\mathrm{ch}}$}}}%
    \put(0.81,0.78){\color[rgb]{0,0,0}\makebox(0,0)[lb]{\smash{$\displaystyle   \Pi_{\mathrm{tot}}^{\mathrm{dis}}$}}}%
    \put(0.035,0.26){\color[rgb]{0,0,0}\makebox(0,0)[lb]{\smash{\begin{rotate}{90}$\displaystyle  \Pi_{\mathrm{rev}}^{\mathrm{ch}}~\big[$J$\big]$\end{rotate}}}}%
    \put(0.4,0.485){\color[rgb]{0,0,0}\makebox(0,0)[lb]{\smash{$\displaystyle \frac{ \Omega_{\mathrm{ent}}}{\Omega}$}}}%
  \end{picture}%
  }
\caption{\label{fig4}%
(a) The total heat production during charging $\Pi^{\mathrm{\mathrm{ch}}}_{\mathrm{tot}}$ (circles) and discharging $\Pi^{\mathrm{\mathrm{dis}}}_{\mathrm{tot}}$ (triangles) [see Eqs.~(\ref{eq:heatcharging}) and~(\ref{eq:heatdischarging})] versus the cell voltage $\Psi$ at $\rhos=1, 10, 100, 1000$ mM.
(b) The reversible heat $\Pi_{\mathrm{rev}}^{\mathrm{\mathrm{ch}}}$ [Eq.~(\ref{eq:revheat})] versus potential was measured at $\rhos=1$~M.  Black lines in (a) and (b) indicate quadratic fits through the data as guides to the eye. Identifying $\Pi_{\mathrm{rev}}^{\mathrm{\mathrm{ch}}}=\Omega_{\mathrm{ent}}$, the inset shows the ratio $\Omega_{\mathrm{ent}}/\Omega$ versus potential.} 
\end{figure}
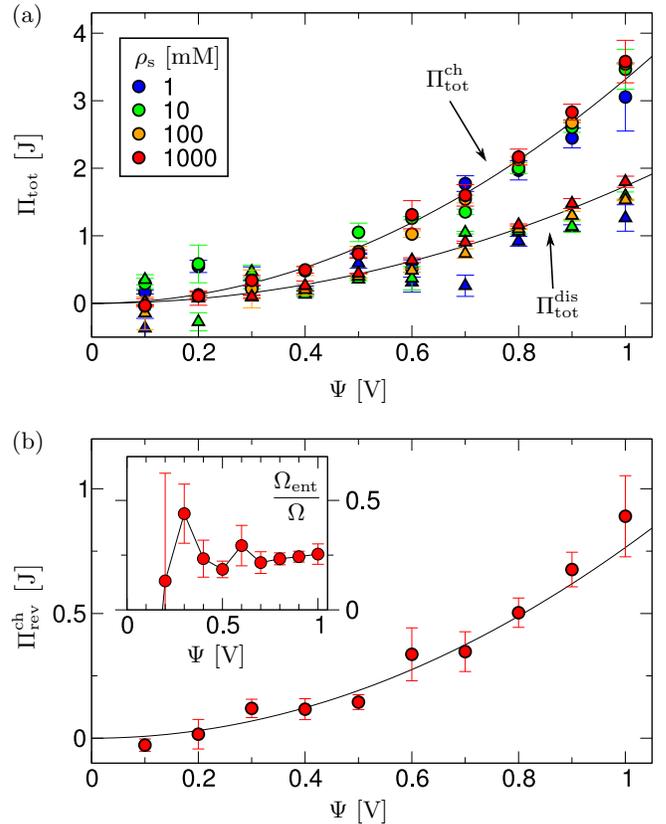

With $K$ at hand, we consider the total heat production during charging, 
\begin{align}\label{eq:heatcharging}
\Pi^{\mathrm{ch}}_{\rm tot}\equiv \Pi_{\rm irr}^{\mathrm{ch}}+\Pi_{\rm rev}^{\mathrm{ch}}=K\int_{t_{0}}^{t_{1}}\Delta T(t) dt,
\end{align} 
and discharging, 
\begin{align}\label{eq:heatdischarging}
\Pi^{\mathrm{dis}}_{\rm tot}\equiv\Pi_{\rm irr}^{\mathrm{dis}}+\Pi_{\rm rev}^{\mathrm{dis}}=K\int_{t_{1}}^{t_{2}}\Delta T(t) dt,
\end{align} 
separately. Using $K=K^{\textrm{h.e.}}$, we plot $\Pi^{\mathrm{ch}}_{\rm tot}$ and $\Pi^{\mathrm{dis}}_{\rm tot}$ as a function of $\Psi$ for different salt concentrations in Fig.~\ref{fig4}(a). We observe that the salt concentration has only a minor effect; results at $\rhos=1$~mM did not differ very much from those at $\rhos=1$~M. 
The reversible heat production is now determined via
\begin{align}\label{eq:revheat}
\Pi^{\mathrm{ch}}_{\rm rev}=\frac{\Pi^{\mathrm{ch}}_{\rm tot}-\Pi^{\mathrm{dis}}_{\rm tot}}{2},
\end{align} 
where we used that $\Pi^{\mathrm{ch}}_{\rm rev}=-\Pi^{\mathrm{dis}}_{\rm rev}$ (by definition) and that $\Pi_{\rm irr}^{\mathrm{dis}}=\Pi_{\rm irr}^{\mathrm{ch}}$  (see Appendix~\ref{sec:carboncalibration}). 
The potential dependence of $\Pi^{\mathrm{ch}}_{\rm rev}$ is shown in Fig.~\ref{fig4}(b). 
Since $\Pi_{\rm rev}^{\mathrm{ch}}$ is an equilibrium property of the EDL, it should be independent of the thermal boundary condition (insulated or not), so we could find the temperature rise upon adiabatic charging from Eq.~(\ref{eq:heatequation}) by setting $K=0$, leading at $\Psi=1$ V to $\Delta T_{\rm adiab}=\Pi_{\rm rev}^{\mathrm{ch}}/\mathbb{C}_{p}=0.082\pm0.016$ K. 
It is reassuring to see that the coulometric and calorimetric experiments provide comparable predictions for $\Delta T_{\rm adiab}$ via two completely independent routes.

Identifying $\Pi_{\mathrm{rev}}^{\mathrm{\mathrm{ch}}}=\Omega_{\mathrm{ent}}$, we have experimental access to the ratio $\Omega_{\mathrm{ent}}/\Omega$, where, for isothermal charging processes, the grand potential  $\Omega=\int\Psi dQ$ equals (minus) the electric work required to charge the cell, which we find by integrating an interpolation through the capacitance data of Fig.~\ref{fig1}(b).
This ratio ---shown in the inset of Fig.~\ref{fig4}(b)--- is approximately $25\%$ at high potentials, while its behavior at low potential remains uncertain. 
This finding can be compared (qualitatively) to the predictions of different EDL models. In particular, Gouy-Chapman theory predicts $\Omega_{\mathrm{ent}}/\Omega=1/2$ at low potentials, while for high potentials this ratio approaches unity \cite{overbeek1990role}. This finding for pointlike ions is in stark contrast to later theoretical work that included finite ionic size \cite{kralj1996simple}. There, it was reported that the electrostatic energy $\Omega_{\mathrm{el}}$ gains in relative importance at higher potentials, at the expense of a decrease in the importance of~$\Omega_{\mathrm{ent}}$. Hence, the EDL models that include ionic volume are in qualitative agreement with our experimental findings. 

In this Letter on temperature effects in electric double layer capacitors  we presented two different temperature-dependent experiments involving water-immersed porous carbon electrodes. Both experiments point toward exothermic heating (heat flowing out of the cell) during quasistatic EDL formation and the opposite effect during quasistatic discharging. 
We moreover presented the first experimental measurement of the entropic contribution to the grand potential energy cost of electric double layer formation. At high electrostatic potentials, this term constitutes approximately $25\%$ of the total grand potential energy. The accuracy of this prediction could be improved in future work by building an even more sensitive calorimetric cell. While this first study considered only NaCl in water, future work could also look at the effect of ionic valency and size. The proposed calorimetric method is useful for distinguishing or ruling out theories, which is important for understanding the electric double layer in porous carbon. This is of fundamental as well as practical importance, as the potential of these materials for future energy harvesting and storage can hardly be overstated.

\begin{acknowledgments}
Henkjan Siekman and Peter de Graaf are thanked for making the electrochemical cell, and Bonny Kuipers and Stephan Zevenhuizen are thanked for technical support. This work is part of the D-ITP consortium, a program of the Netherlands Organisation for Scientific Research (NWO) that is funded by the Dutch Ministry of Education, Culture and Science (OCW). RvR acknowledges financial support from an NWO-VICI grant.
\end{acknowledgments}

\begin{appendix}

\section{Theory for heat flow, thermal response, and heat production during EDL buildup}
The classical understanding of the EDL is in terms of an ideally polarizable, flat electrode whose charge is screened via a compact, salt-concentration independent Stern or Helmholtz layer within \aa ngstr\"{o}ms of the surface, together with a Gouy-Chapman layer that extends into the solution on the scale of the salt-concentration dependent  Debye length. Although the charging behavior of the water-immersed porous carbon electrodes of the main text is poorly described by Gouy-Chapman theory, 
it does aid the interpretation of the experiments of the main text. In particular, we will derive explicit expressions for the isothermal heat flow $\mathbb{Q}$ (Appendix~\ref{heatflow}) and the reversible heat production $\Pi_{\rm rev}$ (Appendix~\ref{heatproduction}). A Gouy-Chapman approximation for the adiabatic temperature rise $\Delta T_{\rm adiab}$ (Appendix~\ref{sec_adiabatictemperaturerise}) was previously derived in the supplemental material of Ref.~\cite{janssen2017reversible}. To set the stage, we first review the known Gouy-Chapman expressions for the grand potential energy cost of EDL formation.

\subsection{Gouy-Chapman grand potential}
We consider a single electrode of area $A$ and an adjacent 1:1 electrolyte solution with a dielectric constant $\varepsilon$, at temperature $T$ and at a bulk salt concentration $\rhos$. 
At a finite electrode potential $\Psi$, the surface obtains a surface charge density $\sigma=Q/(Ae)$, with $e$ the elementary charge, such that $\sigma$ has the dimension of inverse area. To screen the surface charge, an inhomogeneous charge density profile develops in the electrolyte phase, characterized by the cationic and anionic density profiles $\rho_{+}(z)$ and $\rho_{-}(z)$, respectively. Here, the $z$-coordinate is defined as the direction out of the electrode into the solution, which runs from $z=0$ at the electrode to $z=\infty$ far into the bulk, where the local electrostatic potential $\psi(z)$ is zero and the ion densities take their bulk values $\rho_{+}(\infty)=\rho_{-}(\infty)=\rhos$.
The ionic density profiles can be obtained via classical density functional theory from an auxiliary functional $\Omega_{\mathcal{V}}$
consisting of an ideal-gas term and a mean-field Coulomb interaction term,
\begin{eqnarray}\label{eq:Ftotal}
&& \frac{\Omega_{\mathcal{V}}[\rho_{+},\rho_{-}]}{A\kbt} =  \int_{0}^{\infty}\bigg\{\sum_{\alpha=\pm}\rho_{\alpha}(z) \Big[ \ln\big(\rho_{\alpha}(z)\Lambda_{\alpha}^3\big) - 1-\beta\mu_{\alpha} \Big] \nn
&&\hspace{3.5cm}+ \frac{1}{2}q(z) \phi(z) \bigg\}dz, 
\end{eqnarray}
with $k_{B}$ Boltzmann's constant,  $\Lambda_{\alpha}$ the thermal wavelength, $\mu_{\alpha}$ the ionic chemical potential, $\phi(z)=e \psi(z)/\kbt$ the local dimensionless electrostatic potential, and $q(z)=\sigma\delta(z)+\rho_{+}(z)-\rho_{-}(z)$ the local unit charge density.  
The ion density profiles follow from the Euler-Lagrange equation $\delta \Omega/\delta \rho_{\pm}=0$ and amount to $ \rho_{\pm}(z)=\rhos \exp{[\mp\phi(z)]}$ after setting the chemical potential to $\mu_{\pm}=\kbt\ln \rhos \Lambda_{\pm}^{3}$. Inserting $ \rho_{\pm}(z)$ into the Poisson equation, $\phi''(z) = -4\pi {\lb} q(z)$, results in the Poisson-Boltzmann equation.
The main result of Gouy-Chapman theory is the analytic solution to that equation: 
\begin{align} \label{eq:10}
\phi=4  \tanh^{-1}\left(\exp\left[-\ld^{-1}z\right] \tanh{\frac{\Phi}{4}}\right),
\end{align}
with $\Phi=e\Psi/\kbt$ the dimensionless surface potential. Moreover, $\ld=(8\pi\lb\rhos)^{-1/2}$ is the Debye length, in terms of the Bjerrum length $\lb=e^2/(4\pi \varepsilon_{0} \varepsilon \kbt)$, and the vacuum permittivity $\varepsilon_{0}$. From Eq.~(\ref{eq:10}), various other quantities can be derived, for instance the relation between the surface charge and potential
\begin{equation} \label{eq:gouy_chapman}
\sigma=\bar{\sigma} \sinh\frac{\Phi}{2},
\end{equation}
with $\bar{\sigma}=4\rhos\ld$ the crossover surface charge density.

According to density functional theory, we find the grand potential $\Omega(T,V_{\mathrm{el}},\mu_{+},\mu_{-},Q)= \Omega_{\mathcal{V}}[\rho_{+},\rho_{-}]$ (with $V_{\mathrm{el}}$ the electrolyte volume) by evaluating the auxiliary functional at the equilibrium ion concentration profiles. Within Gouy-Chapman theory this yields
\begin{widetext}
\begin{align}\label{eq:equilibriumgrandpotential}
 \Omega^{\mathrm{GC}} &= \underbrace{2 A \rhos\kbt\int_{0}^{\infty}dz\left\{ \phi(z) \sinh \phi(z)-\cosh\phi(z)+1\right\}}_{ \displaystyle\equiv  \Omega^{\mathrm{GC}}_{\mathrm{ent}}}+  \underbrace{\frac{A\kbt}{2}\int_{0}^{\infty}dz\left[ q(z)\phi(z)\right]}_{  \displaystyle\equiv  \Omega^{\mathrm{GC}}_{\mathrm{el}}},
\end{align}
\end{widetext}
where we subtracted the bulk grand potential $-pV_{\mathrm{el}}$ of the electrolyte with osmotic pressure $p=2\rhos\kbt$. The term $\Omega^{\mathrm{GC}}_{\mathrm{ent}}$, the integrand of which goes to zero in the bulk, is associated with the excess entropy of the double layer. Both integrals in Eq.~(\ref{eq:equilibriumgrandpotential}) were solved in Ref.~\cite{overbeek1990role} by employing the Gouy-Chapman solution Eq.~(\ref{eq:10}). The resulting expression for the entropic contribution reads
\begin{align}\label{eq:omegaent}
\Omega_{\mathrm{ent}}^{\mathrm{GC}}&=  A\bar{\sigma} \kbt\left[\Phi \sinh\frac{\Phi}{2} -3 \cosh\frac{\Phi}{2}+3\right],
\end{align}
and, likewise
\begin{align}\label{eq:omegael}
\Omega^{\mathrm{GC}}_{\mathrm{el}}&=A\bar{\sigma} \kbt\left[\cosh{\frac{\Phi}{2}}-1\right],
\end{align}
was found for the electrostatic contribution.

\subsection{Isothermal heat flow}\label{heatflow}
The amount of heat that is required to flow into a capacitor during isothermal ($dT=0$) charging at temperature $T$ from an uncharged state to a charged state with charge $Q$ and potential $\Psi$ was derived in Ref.~\cite{hartel2015heat}:
\begin{equation}\label{eq:thermodynamicidentity}
\mathbb{Q}= -T\int_{0}^{Q}\left(\frac{\partial \Psi(Q',T)}{\partial T}\right)_{Q'}dQ'.
\end{equation}
The coulometric experiment determines temperature dependence of the equilibrium charge, $\left(\partial Q/\partial T\right)_{\Psi}$, which is related to the temperature dependence of the equilibrium potential via a cyclic reciprocity relation  
\begin{equation}\label{eq_experimentalthermalvoltagerise}
\left(\frac{\partial \Psi}{\partial T}\right)_{Q}=-\left(\frac{\partial \Psi}{\partial Q}\right)_{T}\left(\frac{\partial Q}{\partial T}\right)_{\Psi}.
\end{equation}
Hence, we can equivalently express the isothermal heat flow as
\begin{equation}\label{eq:isothermalheatflow2}
\mathbb{Q}= T\int_{0}^{\Psi}\left(\frac{\partial Q(\Psi',T)}{\partial T}\right)_{\Psi'}d \Psi'.
\end{equation}
In the coulometric experiment of the main text we found  $\left(\partial Q/\partial T\right)_{\Psi}=\alpha\Psi$, with $\alpha=-4.0\pm0.1$~mC~V$^{-1}$~K$^{-1}$. Inserting this into the above equation we find $\mathbb{Q}=\alpha T\Psi^2/2$: the result stated in the main text.
 
Within Gouy-Chapman the heat flow reads [after inserting Eq.~(\ref{eq:gouy_chapman})]
\begin{align}\label{eq:gouychapmanisothermalheat}
\mathbb{Q}^{\rm GC}
&=-2A\kbt\int_{0}^{\sigma }\frac{\partial }{\partial T}\left[T\sinh^{-1} \frac{\sigma'}{\bar{\sigma}}\right]d\sigma'\nn
&= -2A\kbt\bigg[\sigma\sinh^{-1} \frac{\sigma}{\bar{\sigma}}-\sqrt{\bar{\sigma}^{2}+\sigma^{2}}+\bar{\sigma}\bigg],
\end{align}
where we ignored the temperature dependence of $\bar{\sigma}$.
In terms of the dimensionless surface potential $\Phi$, and comparing to Eqs.~(\ref{eq:omegaent}) and (\ref{eq:omegael}), we find
\begin{align}\label{eq:}
\mathbb{Q}^{\rm GC}=& -A\bar{\sigma}\kbt \bigg[2-2\cosh\frac{\Phi}{2}+\Phi\sinh\frac{\Phi}{2}\bigg]\nn
=&-\Omega_{\mathrm{ent}}^{\mathrm{GC}}.
\end{align}
Thermodynamically, this equivalence is understood as follows. The differential $d\Omega=-SdT+\Psi dQ$ of the grand potential (at fixed $\mu_{\pm}$ and $V_{\mathrm{el}}$) of the electric double layer capacitor
implies that the isothermal grand potential change during charging is equal to (minus) the performed work, $ \Omega=-W=\int\Psi dQ$.
Meanwhile, the isothermal heat Eq.~(\ref{eq:thermodynamicidentity}) is equal and opposite, $\mathbb{Q}=-\int \Psi dQ$, {\it if} the charging behavior is such that $\left(\partial \Psi/\partial T\right)_{Q}=\Psi/T$, which is precisely what we assumed in this section by neglecting $\partial \bar{\sigma}/\partial T$ in Eq.~(\ref{eq:gouychapmanisothermalheat}). This implies moreover that the internal energy change is zero for the Gouy-Chapman model. In general, more advanced EDL models will have a nonzero internal energy change upon charging, in which case the heat flow is not equal to the grand potential  change. 

\subsection{Adiabatic temperature rise}\label{sec_adiabatictemperaturerise}
Slowly charging a thermally insulated electric double layer capacitor leads to a temperature rise $\Delta T_{\rm adiab}$, which is the capacitive analogue of the temperature increase upon compressing a gas in a thermally insulated container \cite{Janssen:2014aa}. Here, $\Delta T_{\rm adiab}$ is found by integrating the relation 
\begin{equation}\label{eq:adiabatic_Tincrease}
dT=\frac{T}{\mathbb{C}_{p}}\left(\frac{\partial Q(\Psi,T)}{\partial T}\right)_{\Psi}d \Psi,
\end{equation}
that follows from the total differential of $S(T,\Psi)$, together with the $dS=0$ condition on isentropic processes.  Using once more $\left(\partial Q/\partial T\right)_{\Psi}=\alpha\Psi$ (with $\alpha=-4.0\pm0.1$~mC~V$^{-1}$~K$^{-1}$) as obtained with the coulometric experiment of the main text, we find a small adiabatic temperature rise,
\begin{equation}
\Delta T_{\rm adiab}=\frac{\alpha T\Psi^2}{2\mathbb{C}_{p}},
\end{equation}
upon increasing the potential from 0 to $\Psi$, with $T$ the initial temperature at the uncharged state.

\subsection{Reversible heat production}\label{heatproduction}
Overbeek pointed out that the work performed  by the electric field on ions in a solution during quasistatic EDL buildup equals the entropic contribution to the grand potential  $\Omega_{\mathrm{ent}}$ \cite{overbeek1990role}. Meanwhile, the source term ${\bf I}\cdot{\bf E}$ appearing in the heat equation~\eqref{eq:microheatequation} is associated precisely with the power delivered locally by the electric field for out of equilibrium settings. This suggests the following relation
\begin{equation} \label{eq:3}
\Pi_{\rm rev}^{\mathrm{ch}}\equiv\lim_{\mathcal{T}\to\infty}\int_{0}^{\mathcal{T}} dt\int_{\mathcal{V}_{\rm el}} d{\bf x}~{\bf I}\cdot{\bf E}=\Omega_{\mathrm{ent}},
\end{equation}
where $\mathcal{T}$ is the duration of the slow charging process.
To test  Eq.~(\ref{eq:3}), we consider a very slow charging process of an electrode from an uncharged state, to some final state at charge $Q$. We study the electrode as described in the introduction of this section, and assume that the instantaneous density profiles are given as in Gouy-Chapman theory [Eq.~(\ref{eq:10})] and moreover that the temperature $T(z,t)$ is homogeneous throughout the electrolyte volume at each time $t$. Under these conditions, the integrand of Eq.~(\ref{eq:3}) simplifies to $IE$, with $E=-\partial_{z}\psi$ the local electric field, and $I=eJ_{\mathrm{\mathrm{ch}}}=e(J_{+}-J_{-})$ the ionic current density (in A m$^{-2}$). 
The time integral in Eq.~(\ref{eq:3}) can be transformed into an integral over the unit surface charge \mbox{density $\sigma$}
\begin{align} \label{eq:4}
\Pi_{\rm rev}^{\mathrm{ch, GC}}=-A\kbt\int_{0}^{\sigma} \left[ \int_0^\infty \frac{J_\mathrm{\mathrm{ch}}}{J_\text{tot}}\> \frac{\partial \phi}{\partial z} \> dz \right] d\sigma',
\end{align}
where we implemented 
\begin{equation} \label{eq:5}
\frac{\partial \sigma}{\partial t}=J_\text{tot},
\end{equation}
with $J_{\text{tot}}=J_{\mathrm{\mathrm{ch}}}+J_{\mathrm{M}}$ the total current density, which is equal to the electronic current entering the electrode. Here, $J_{\text{tot}}$, invariant with position in a Cartesian one-dimensional geometry, includes the Maxwell or displacement current $J_{\mathrm{M}}=-\varepsilon_{0}\varepsilon\partial_{t}\partial_{z}\psi$, which is proportional to the variation of the electric field in time \cite{van2010diffuse}.  As the Maxwell current vanishes in the charge neutral bulk, $J_\text{tot}$ is equal to the ionic current $J_{\mathrm{ch}}$ far from the electrode. For capacitive charging of the electrode, the ionic current $J_{\mathrm{ch}}$ is zero at the electrode surface and increases with $z$ to reach $J_\text{tot}$ outside the EDL. Thus, in general $J_{\mathrm{ch}}$ is a fraction of $J_\text{tot}$, i.e., $0<J_{\mathrm{ch}}/J_\text{tot}<1$. 
The ionic current density $J_{\mathrm{ch}}$ relates to the local ionic charge density $q$ according to the continuity equation
\begin{equation} \label{eq:6}
\frac{\partial J_{\mathrm{ch}}}{\partial z}=-\frac{\partial q}{\partial t}.
\end{equation}
Within Gouy-Chapman  theory $q=-2 \rhos \sinh{\phi}$ and thus Eqs.~(\ref{eq:5}) and (\ref{eq:6}) yield
\begin{equation} \label{eq:9}
 \frac{\partial}{\partial z}\left(\frac{J_{\mathrm{ch}}}{J_\text{tot}}\right)=2 \rhos \cosh{\phi}\frac{\partial \phi}{\partial \Phi}\frac{\partial \Phi}{\partial \sigma}. 
\end{equation}
We perform one partial integration in the bracketed integral in Eq.~(\ref{eq:4}) and insert Eq.~(\ref{eq:9}), such that
\begin{align} \label{eq:13}
\frac{\Pi_{\rm rev}^{\mathrm{ch, GC}}}{A\kbt}&=-\int_{0}^{\sigma} \left[\int_0^\infty \frac{\partial }{\partial z}\left(\frac{J_{\mathrm{ch}}}{J_\text{tot}}\right) \phi \> dz \right] d\sigma' \nn
&=-2 \rhos \int_{0}^{\sigma} \left[\int_0^\infty  \phi \cosh{\phi}\frac{\partial \phi}{\partial \Phi}\frac{\partial \Phi}{\partial \sigma}  \> dz \right] d\sigma' \nn
&=-2 \rhos \int_{0}^{\Phi} \left[\int_0^\infty  \phi \cosh{\phi}\frac{\partial \phi}{\partial \Phi}  \> dz \right] d\Phi' \nn
&=-2 \rhos \int_{0}^{\Phi} \left[\int_{\Phi'}^0  \phi \cosh{\phi}\frac{\partial \phi}{\partial \Phi}  \frac{\partial z}{\partial \phi} \> d\phi \right] d\Phi' ,
\end{align}
where in the second to last step we took $\left(\partial \Phi/\partial \sigma\right)$ out of the spatial integral since it does not depend on~$z$. 
The integrand of Eq.~(\ref{eq:13}) can be simplified considerably, in particular by converting the relation $\phi(\Phi,z)$ [Eq.~(\ref{eq:10})] into $\Phi(\phi,z)$. This gives the following term that can be taken out of $\phi-$integral in Eq.~(\ref{eq:13}),
\begin{align}
\frac{\partial \phi}{\partial \Phi}  \frac{\partial z}{\partial \phi}
&=\left(\frac{\partial \Phi}{\partial z}\right)^{-1}
=\frac{\ld}{4}\frac{\left[1-\exp\left[2z/\ld\right] \tanh^{2}\frac{\phi}{4}\right]}{\exp\left[z/\ld\right] \tanh\frac{\phi}{4}}\nn
&=\frac{\ld}{4}\frac{\left[1-\tanh^{2}\frac{\Phi}{4}\right]}{\tanh\frac{\Phi}{4}}
=\frac{\ld}{2\sinh\frac{\Phi}{2}},
\end{align}
where, going to second line, we eliminated $z$ via Eq.~(\ref{eq:10}). We can now evaluate Eq.~(\ref{eq:13}), 
\begin{align}
\frac{\Pi_{\rm rev}^{\mathrm{ch, GC}}}{A\kbt}&=-\rhos\ld \int_{0}^{\Phi}\frac{1}{\sinh\frac{\Phi'}{2}}\left[\int_{\Phi'}^{0}\phi\cosh\phi \> d\phi\right]  \> d\Phi'\nn
&=-\rhos \ld\int_{0}^{\Phi}\frac{\cosh\Phi'-\Phi'\sinh\Phi' -1}{\sinh\frac{\Phi'}{2}}d\Phi'\nn
&=-2\rhos \ld\int_{0}^{\Phi}\left[\sinh\frac{\Phi'}{2}-\Phi'\cosh\frac{\Phi'}{2}\right] d\Phi'\nn
&=\bar{\sigma} \left[3-3\cosh\frac{\Phi}{2}+\Phi\sinh\frac{\Phi}{2}\right] .
\end{align}
Comparing with Eq.~(\ref{eq:omegaent}), we see indeed that \mbox{$\Pi_{\rm rev}^{\mathrm{ch, GC}}=\Omega_{\mathrm{ent}}^{\mathrm{GC}}$}: precisely the hypothesis we set out to test.

\section{The heat transfer coefficient $K$}\label{sec:calibration}
The heat transfer coefficient $K$ appearing in Eq.~\eqref{eq:heatequation} was determined by two calibration methods. 

\subsection{Calibration with heating elements}\label{sec:alternativecalorimetry}
\begin{figure}
\def\svgwidth{0.45\textwidth}{
  \providecommand\color[2][]{%
    \errmessage{(Inkscape) Color is used for the text in Inkscape, but the package 'color.sty' is not loaded}%
    \renewcommand\color[2][]{}%
  }%
  \providecommand\transparent[1]{%
    \errmessage{(Inkscape) Transparency is used (non-zero) for the text in Inkscape, but the package 'transparent.sty' is not loaded}%
    \renewcommand\transparent[1]{}%
  }%
  \providecommand\rotatebox[2]{#2}%
  \ifx\svgwidth\undefined%
    \setlength{\unitlength}{515.84695303bp}%
    \ifx\svgscale\undefined%
      \relax%
    \else%
      \setlength{\unitlength}{\unitlength * \real{\svgscale}}%
    \fi%
  \else%
    \setlength{\unitlength}{\svgwidth}%
  \fi%
  \global\let\svgwidth\undefined%
  \global\let\svgscale\undefined%
  \makeatother%
  \begin{picture}(1,0.83)%
    \put(0,0.05){\includegraphics[width=\unitlength]{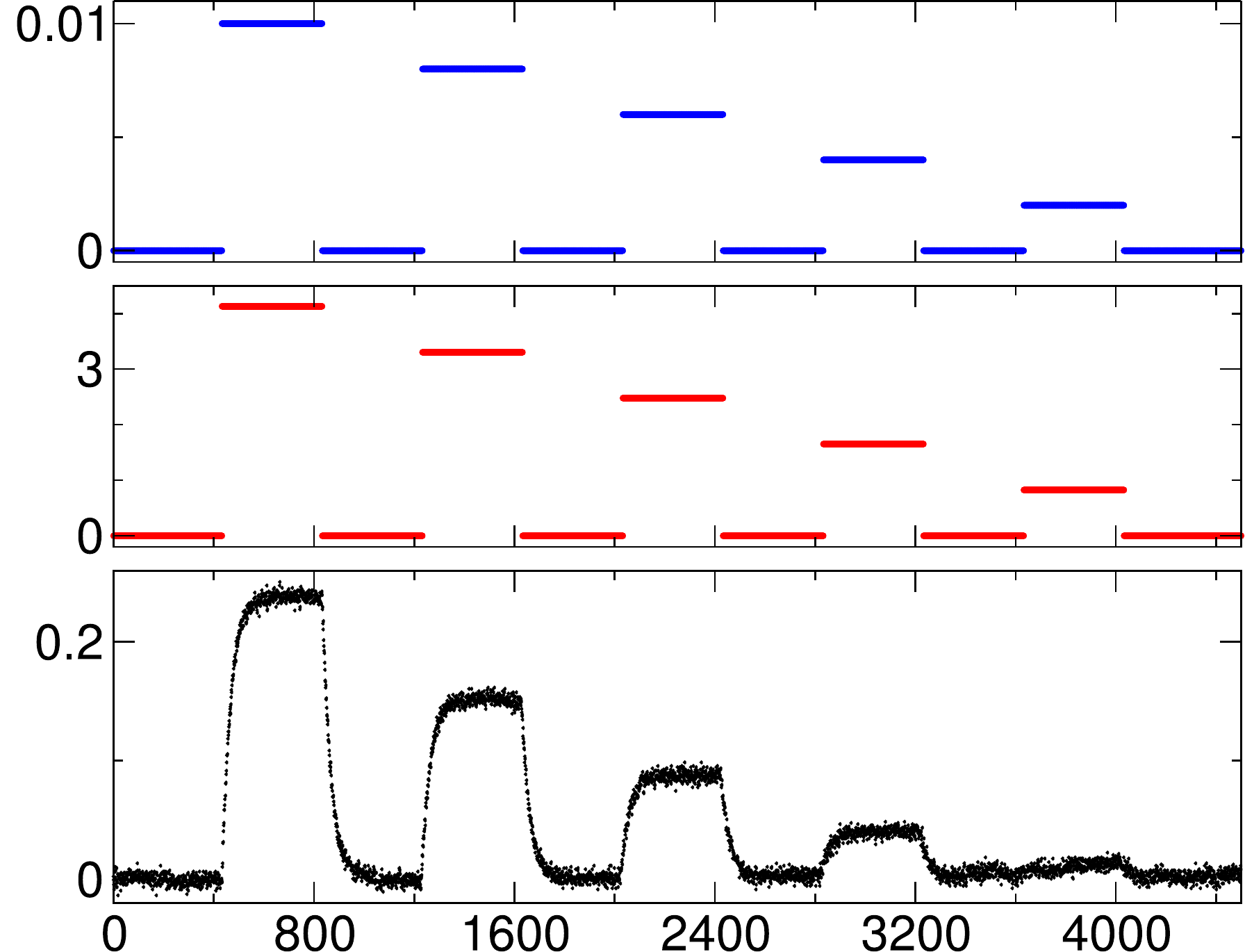}}%
    \put(0.92,0.77){\color[rgb]{0,0,0}\makebox(0,0)[lb]{\smash{(a)}}}%
    \put(0.92,0.54){\color[rgb]{0,0,0}\makebox(0,0)[lb]{\smash{(b)}}}%
    \put(0.92,0.31){\color[rgb]{0,0,0}\makebox(0,0)[lb]{\smash{(c)}}}%
    \put(-0.000,0.43){\color[rgb]{0,0,0}\makebox(0,0)[lb]{\smash{\begin{rotate}{90} $\displaystyle \Psi_{\mathrm{R}}$~[V]\end{rotate}}}}%
    \put(0.00,0.67){\color[rgb]{0,0,0}\makebox(0,0)[lb]{\smash{\begin{rotate}{90} $\displaystyle I$~[A]\end{rotate}}}}%
    \put(0.0,0.15){\color[rgb]{0,0,0}\makebox(0,0)[lb]{\smash{\begin{rotate}{90} $\displaystyle \Delta T$~[K]\end{rotate}}}}%
     \put(0.5,0.0){\color[rgb]{0,0,0}\makebox(0,0)[lb]{\smash{ $t$~[s]}}}%
  \end{picture}%
  }
\caption{\label{fig_sup2}%
In the calibration with the heating elements, the current $I$ (a) was cyclically switched on and off to different values, each lasting 400 s. The voltage $\Psi_{\rm R}$ (b) over the resistive wire, and temperature difference $\Delta T$ (c) between the cell and the thermostatic bath  were measured simultaneously. Shown are data for the air-filled cell configuration.}
\end{figure}
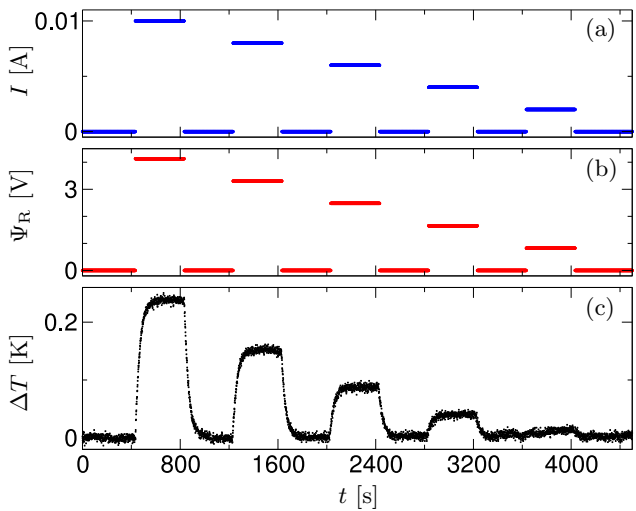

Experiments were carried out with the same cell as described in the main text, but with the electrodes and current collectors replaced by heating elements (positioned at the same respective locations as the electrodes) consisting of resistive wire glued as a flat spiral onto a glass disk (total resistance: 413~$\Omega$). 
Measurements were conducted with the cell filled with water and with the cell filled with air. Depending on the current $I$ [see Fig.~\ref{fig_sup2}(a)] of up to 10~mA, different voltages $\Psi_{\rm R}$ [Fig.~\ref{fig_sup2}(b)] over the resistive wire were measured, and different plateau values $\Delta T_{\rm max}$ of the temperature difference $\Delta T$  [Fig.~\ref{fig_sup2}(c)] between the cell and the thermostatic bath were reached. 

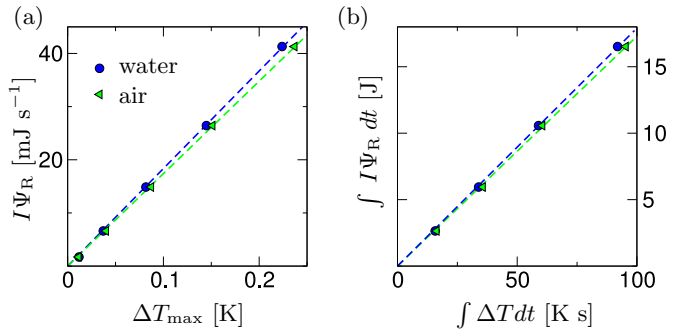
\begin{figure}
\def\svgwidth{0.49\textwidth}{
  \providecommand\color[2][]{%
    \errmessage{(Inkscape) Color is used for the text in Inkscape, but the package 'color.sty' is not loaded}%
    \renewcommand\color[2][]{}%
  }%
  \providecommand\transparent[1]{%
    \errmessage{(Inkscape) Transparency is used (non-zero) for the text in Inkscape, but the package 'transparent.sty' is not loaded}%
    \renewcommand\transparent[1]{}%
  }%
  \providecommand\rotatebox[2]{#2}%
  \ifx\svgwidth\undefined%
    \setlength{\unitlength}{610.97107821bp}%
    \ifx\svgscale\undefined%
      \relax%
    \else%
      \setlength{\unitlength}{\unitlength * \real{\svgscale}}%
    \fi%
  \else%
    \setlength{\unitlength}{\svgwidth}%
  \fi%
  \global\let\svgwidth\undefined%
  \global\let\svgscale\undefined%
  \makeatother%
  \begin{picture}(1,0.46)%
    \put(0,0.032){\includegraphics[width=\unitlength]{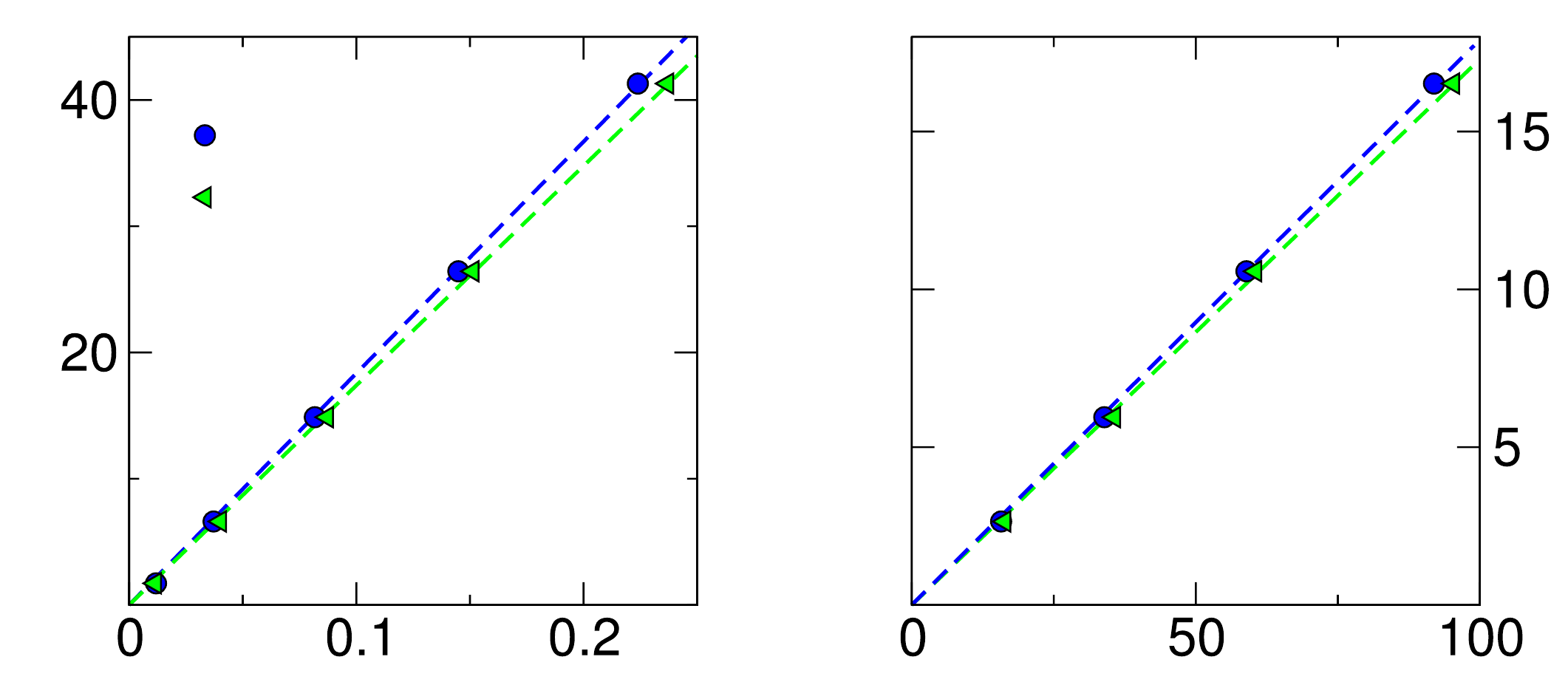}}%
    \put(0.18,0.00){\color[rgb]{0,0,0}\makebox(0,0)[lb]{\smash{$\Delta T_{\rm max}$ [K]}}}%
    \put(0.67,0.00){\color[rgb]{0,0,0}\makebox(0,0)[lb]{\smash{$\int\Delta T dt$ [K s]}}}%
    \put(0.55,0.17){\color[rgb]{0,0,0}\makebox(0,0)[lb]{\smash{\begin{rotate}{90} $\int\>I\Psi_{\mathrm{R}}\>dt$ [J]\end{rotate}}}}%
    \put(0.025,0.15){\color[rgb]{0,0,0}\makebox(0,0)[lb]{\smash{\begin{rotate}{90} $I\Psi_{\mathrm{R}}$ [mJ s$^{-1}$]\end{rotate}}}}%
    \put(0.16,0.375){\color[rgb]{0,0,0}\makebox(0,0)[lb]{\smash{water}}}%
    \put(0.16,0.33){\color[rgb]{0,0,0}\makebox(0,0)[lb]{\smash{air}}}%
      \put(0.0,0.45){\color[rgb]{0,0,0}\makebox(0,0)[lb]{\smash{(a)}}}%
    \put(0.49,0.45){\color[rgb]{0,0,0}\makebox(0,0)[lb]{\smash{(b)}}}%
  \end{picture}%
  }
\caption{\label{fig_sup3}%
The heat transfer coefficient $K$ is determined both from the plateau value $\Delta T_{\rm max}$ (a), as well as from the time-integrated temperature peaks $\int\Delta T dt$  (b), both for a water-filled cell (blue circles) and an air-filled cell (green triangles). Dotted lines indicate linear fits through these data.}
\end{figure}

The occurrence of $\Delta T$-plateaus indicates that a steady state is reached where Joule heating is balanced by heat flowing out the cell. From Eq.~\eqref{eq:heatequation} it then follows that the heat transfer coefficient $K$ can be determined as the slope of a $\Delta T_{\rm max}$-versus-$I\Psi_{\mathrm{R}}$ plot. In Fig.~\ref{fig_sup3}(a) we see that $\Delta T_{\rm max}$ is proportional to the heating power $I\Psi_{\mathrm{R}}$. We find $K = 0.174\pm0.001$~J~s$^{-1}$~K$^{-1}$ and $K = 0.184\pm0.001$~J~s$^{-1}$~K$^{-1}$ for the air-filled and water-filled cell, respectively.

Taking a time integral over Eq.~\eqref{eq:heatequation}, its left hand side vanishes for any time interval $t_{a}<t<t_{b}$ which starts ($t_{a}$) and ends ($t_{b}$) at $\Delta T=\SI{0}{\degreeCelsius}$. The heat transfer coefficient can then alternatively be determined from the slope of a plot of $\int_{t_{a}}^{t_{b}}\>I\Psi_{\mathrm{R}}\>dt$ versus $\int_{t_{a}}^{t_{b}}\>\Delta T\>dt$ [Fig.~\ref{fig_sup3}(b)]. 
We found $K = 0.173\pm0.001$~J~s$^{-1}$~K$^{-1}$ for the air-filled cell and $K = 0.179\pm0.001$~J~s$^{-1}$~K$^{-1}$ for the water-filled cell.
Clearly, these values agree well with those determined from Fig.~\ref{fig_sup3}(a); we think that the second method is more reliable because it does not require a determination of the time at which $\Delta T_{\rm max}$ is reached.

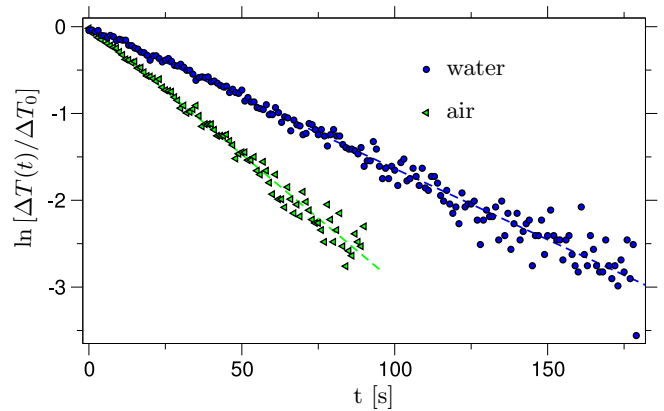
\begin{figure}
\def\svgwidth{0.46\textwidth}
  \providecommand\color[2][]{%
    \errmessage{(Inkscape) Color is used for the text in Inkscape, but the package 'color.sty' is not loaded}%
    \renewcommand\color[2][]{}%
  }%
  \providecommand\transparent[1]{%
    \errmessage{(Inkscape) Transparency is used (non-zero) for the text in Inkscape, but the package 'transparent.sty' is not loaded}%
    \renewcommand\transparent[1]{}%
  }%
  \providecommand\rotatebox[2]{#2}%
  \ifx\svgwidth\undefined%
    \setlength{\unitlength}{543.16897759bp}%
    \ifx\svgscale\undefined%
      \relax%
    \else%
      \setlength{\unitlength}{\unitlength * \real{\svgscale}}%
    \fi%
  \else%
    \setlength{\unitlength}{\svgwidth}%
  \fi%
  \global\let\svgwidth\undefined%
  \global\let\svgscale\undefined%
  \makeatother%
  \begin{picture}(1,0.63)%
    \put(0.04,0.02){\includegraphics[width=\unitlength]{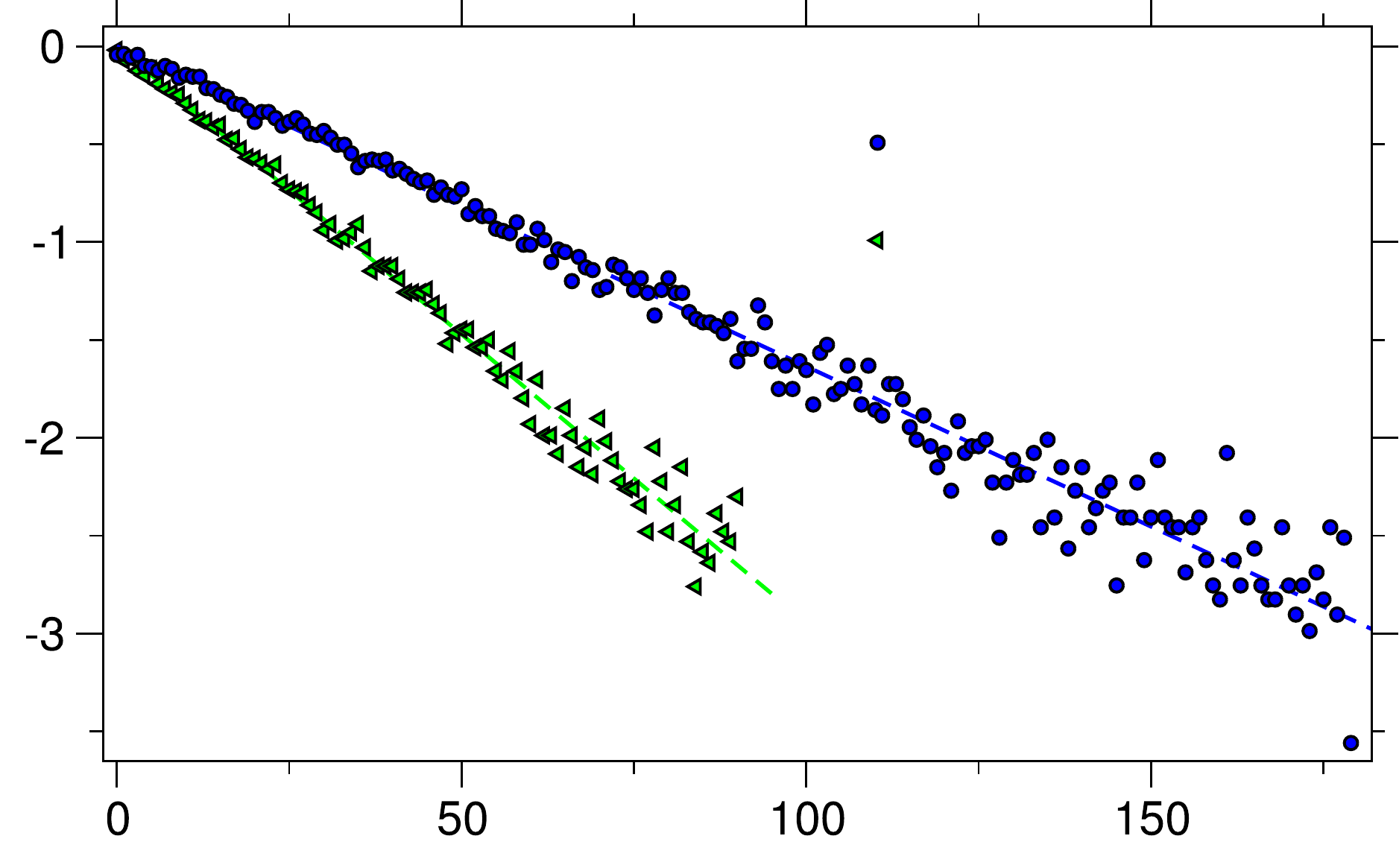}}%
    \put(0.7,0.53){\color[rgb]{0,0,0}\makebox(0,0)[lb]{\smash{water}}}%
    \put(0.7,0.46){\color[rgb]{0,0,0}\makebox(0,0)[lb]{\smash{air}}}%
    \put(0.03,0.25){\color[rgb]{0,0,0}\rotatebox{89.7025396}{\makebox(0,0)[lb]{\smash{$\ln\left[\Delta T(t)/\Delta T_{0}\right]$}}}}%
    \put(0.55,0.00){\color[rgb]{0,0,0}\makebox(0,0)[lb]{\smash{t [s]}}}%
  \end{picture}%
\caption{\label{fig_sup2b}%
The heat capacity $\mathbb{C}_{p}$ of the cell is determined from the exponential decay of  $\Delta T$ after switching off the current $I$. Dotted lines indicate linear fits through these data.}
\end{figure}

Once the applied current was switched off, $\Delta T$ decreased exponentially with time [see Figs.~\ref{fig_sup2}(c) and \ref{fig_sup2b}]; from Eq.~\eqref{eq:heatequation}  follows $\Delta T(t) = \Delta T_{0} \exp(-t/\tau)$, where $\Delta T_{0}$ is the starting value of $\Delta T$, $t$ is the time that passed since the current was switched off, and $\tau = \mathbb{C}_{p}/K$ is the thermal time constant, with $\mathbb{C}_{p}$ the heat capacity of the cell. From the measured time scales $\tau = 34.0\pm0.2$~s for the air-filled cell and $\tau = 61.1\pm0.3$~s for the water-filled cell, we calculate heat capacities of $\mathbb{C}_{p} = K\tau = 6.1\pm0.1$~J~K$^{-1}$ and $10.9\pm0.1$~J~K$^{-1}$, respectively. The difference between the two $\mathbb{C}_{p}$ values agrees within~5\% with a cell volume of approximately 1.1 mL and the specific heat capacity of water, 4.2~J~K$^{-1}$~g$^{-1}$. This consistency supports the validity of the heat quantities obtained. 

In the main text we determined the adiabatic temperature rise $\Delta T_{\rm adiab}$ associated with charging of thermally insulated porous carbon electrodes via two independent routes. We note that both routes used $\mathbb{C}_{p}=10.9\pm0.1$~J~K$^{-1}$ as determined above, that is, with the heating elements. This means that the true heat capacity of the cell might differ slightly since it contains carbon electrodes instead of resistive wires.  

\subsection{Calibration with porous carbon}\label{sec:carboncalibration}
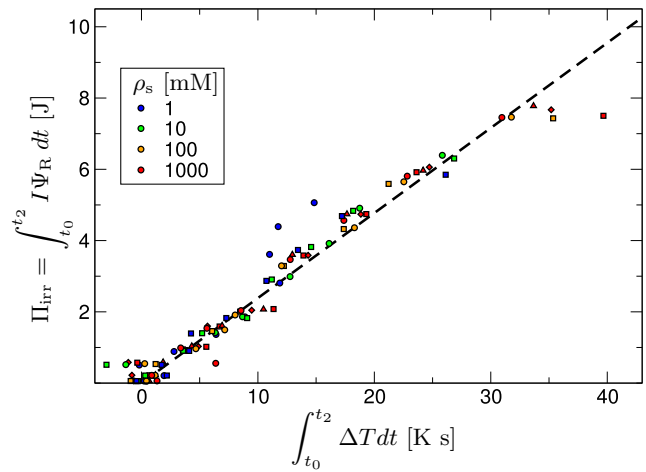
\begin{figure}
\def\svgwidth{0.45\textwidth}{ 
  \providecommand\color[2][]{%
    \errmessage{(Inkscape) Color is used for the text in Inkscape, but the package 'color.sty' is not loaded}%
    \renewcommand\color[2][]{}%
  }%
  \providecommand\transparent[1]{%
    \errmessage{(Inkscape) Transparency is used (non-zero) for the text in Inkscape, but the package 'transparent.sty' is not loaded}%
    \renewcommand\transparent[1]{}%
  }%
  \providecommand\rotatebox[2]{#2}%
  \ifx\svgwidth\undefined%
    \setlength{\unitlength}{543.40262003bp}%
    \ifx\svgscale\undefined%
      \relax%
    \else%
      \setlength{\unitlength}{\unitlength * \real{\svgscale}}%
    \fi%
  \else%
    \setlength{\unitlength}{\svgwidth}%
  \fi%
  \global\let\svgwidth\undefined%
  \global\let\svgscale\undefined%
  \makeatother%
  \begin{picture}(1,0.73)%
    \put(-0.02,0.04){\includegraphics[width=\unitlength]{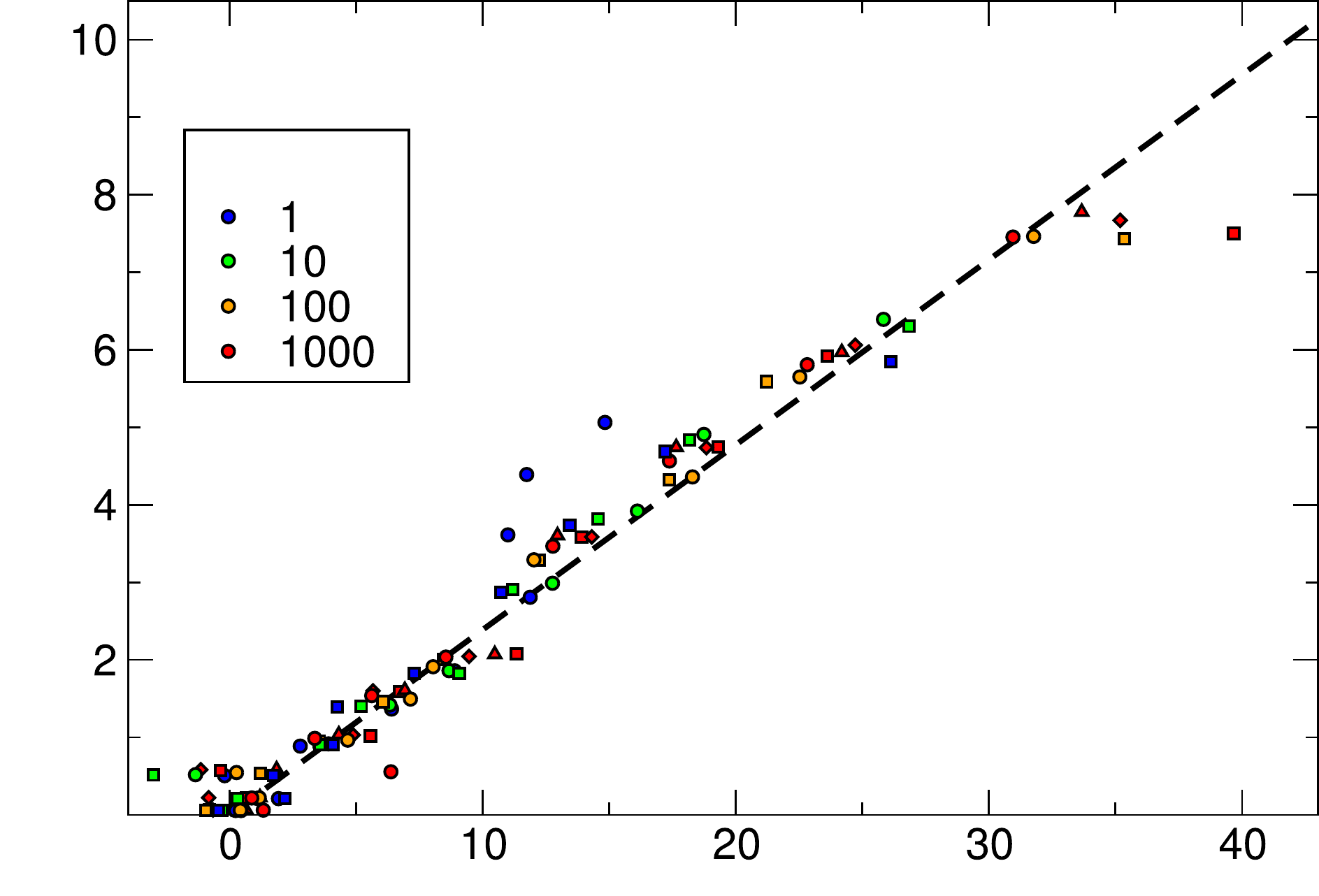}}%
    \put(-0.0000,0.2){\color[rgb]{0,0,0}\makebox(0,0)[lb]{\smash{\begin{rotate}{90}$\displaystyle \Pi_{\mathrm{irr}}=\int_{t_{0}}^{t_{2}} I\Psi_{\mathrm{R}}\>dt$ [J]\end{rotate}}}}%
    \put(0.4,0.0){\color[rgb]{0,0,0}\makebox(0,0)[lb]{\smash{$\displaystyle\int_{t_{0}}^{t_{2}}\Delta T dt$~[K~s]}}}%
    \put(0.14,0.585){\color[rgb]{0,0,0}\makebox(0,0)[lb]{\smash{$\rhos$ [mM]}}}%
  \end{picture}%
}
\caption{\label{fig_sup1}%
In the calibration with porous carbon electrodes, the  heat transfer coefficient $K^{\mathrm{c.e.}} = 0.239\pm 0.003$~J K$^{-1}$~s$^{-1}$ is determined from the slope of the parametric plot of time-integrated $\Delta T$ peaks against the Joule heat $\Pi_{\mathrm{irr}}$ for several salt concentrations, indicated with different colors (see legend). Different symbols belong to different duplicate measurements at various cell potentials up to 1 V.}
\end{figure}

The heat transfer coefficient $K$ can also be determined using the Joule heat that is generated during EDL formation in experiments with porous carbon electrodes. Employing the same arguments as presented in Appendix~\ref{sec:alternativecalorimetry}, a time integral over Eq.~\eqref{eq:heatequation} will leave its left hand side zero for any time interval that starts  and ends at $\Delta T=\SI{0}{\degreeCelsius}$.
In particular, for a full cycle of charging and discharging as shown in Fig.~\ref{fig3}  we have
\begin{align}\label{eq:Kdetermination}
K\int_{t_{0}}^{t_{2}}\Delta T dt&=\Pi_{\rm irr}^{\mathrm{ch}}+\Pi_{\rm irr}^{\mathrm{dis}}\equiv \Pi_{\rm irr},
\end{align} 
where we used Eq.~\eqref{eq:productionsplitting} and the fact that the reversible heat $\Pi_{\rm rev}^{\mathrm{dis}}$ that is produced upon discharging will exactly cancel the reversible heat $\Pi_{\rm rev}^{\mathrm{ch}}$ produced upon charging ($\Pi_{\rm rev}^{\mathrm{ch}}=-\Pi_{\rm rev}^{\mathrm{dis}}$). 
Thus, the only heat production appearing in Eq.~(\ref{eq:Kdetermination}) is Joule heat dissipated from the resistive parts of the system, 
which is exothermic regardless of the opposite signs of voltages and currents during charging and discharging.
To calculate this Joule heat, we time-integrate the dissipated power $\dot{\Pi}_{\rm irr}=I(t)\Psi_{\mathrm{R}}(t)$ over the two time intervals. 
Here, $\Psi_{\mathrm{R}}(t)=\Psi-\Psi_{\mathrm{EDL}}(t)$ is the voltage drop across the resistive elements that dissipate heat, 
which we can access experimentally if we assume a voltage drop $\Psi_{\mathrm{EDL}}(t) = Q(t)/C$ over the electric double layer (EDL), with $Q(t)$ the charge in the EDL and the capacitance $C=\lim_{t\to\infty}Q(t)/\Psi$ is assumed constant. 
Interestingly, we find that equal amounts of Joule heat are produced during charging ($\Pi_{\rm irr}^{\mathrm{ch}}$) and discharging ($\Pi_{\rm irr}^{\mathrm{dis}}$), despite the difference in $I(t)$ and $\Psi_{\mathrm{R}}(t)$ during these processes. The heat transfer coefficient $K$ can now be determined from Fig.~\ref{fig_sup1}, where we plot $\Pi_{\rm irr}=\int_{t_{0}}^{t_{2}} I(t)\Psi_{\mathrm{R}}(t)dt$ versus $\int_{t_{0}}^{t_{2}}\Delta T dt$, obtained for different salt concentrations $\rhos$ and different voltage steps $\Psi$. A straight line is observed, whose slope corresponds to $K = 0.239\pm 0.003$~J K$^{-1}$~s$^{-1}$. Considering only the data at $\rhos=1$~M, the linear fit has a slope $K = 0.232\pm 0.005$~J~K$^{-1}$~s$^{-1}$, hence, $K$ depends very weakly on the salt concentration and can be regarded constant. 

\subsection{Discussion}\label{sec:discussion}
In the presence of carbon electrodes (c.e.), the heat transfer coefficient $K^{\mathrm{c.e.}}=0.239$~J~K$^{-1}$~s$^{-1}$ is a bit higher than in the case of the  413~$\Omega$ heating elements (h.e.), where we found $K^{\mathrm{h.e.}} = 0.179$~J~s$^{-1}$~K$^{-1}$. Both experiments have their pros and cons, which we now discuss.

\textbf{Materials:} the calibration with the porous carbon has the benefit that it does not alter the cell setup. The different materials in the experiment with the resistive wires might affect its $K$-value. 

\textbf{Joule heat determination:}
Both calibration methods use the Joule heat produced over a time interval that starts and ends at $\Delta T=\SI{0}{\degreeCelsius}$. While the resistive wire has a precisely-known resistance, allowing for accurate Joule heat determination, the Joule heat in the porous carbon experiments is prone to larger uncertainty, since we assumed a constant capacitance in its determination. Relaxing this assumption is not straightforward without additional assumptions. 

\textbf{Geometry:}
In both calibration methods, reversible heat can be found only after deducing $K$ from \mbox{$\Delta T$-signals}, which is done via Eq.~\eqref{eq:heatequation} with an (approximately) known amount of Joule heat. We note that this equation relies on the assumption that the temperature is uniform throughout the electrochemical cell ($\mathcal{V}_{\rm s}$ and $\mathcal{V}_{\rm el}$), for which the Biot number Bi, comparing heat conduction within the cell to the heat loss to the environment, is commonly used as a benchmark. 
Unfortunately, even when the requirement of small Bi is satisfied (for our cell we estimate Bi~$\approx0.2$, a bit higher than Bi~$\approx0.1$ which is deemed sufficiently low \cite{incropera2007fundamentals}), the assumption of a uniform temperature cannot be correct for far-out-of-equilibrium charging, since heat is generated quickly at specific locations within the cell.

These considerations have the following repercussions on our experiments.
In both calibration methods, the temperature probe was placed halfway between the two electrodes and heating elements, respectively. Meanwhile, the Joule heat is generated at different locations in the two calibration methods. In the heating elements experiment, Joule heat is generated in a resistive wire positioned at the exact same location as the electrodes in the porous carbon experiment; hence, the location of reversible heat production.
Meanwhile, the weak salt-concentration dependence of the total resistance of the electrochemical cell (that was mentioned in the main text) indicates that in the porous carbon experiment, Joule heat is probably produced at the electrical contacts [see Fig.~\ref{fig1}(a)], which is further away from the temperature probe than in the heating elements experiment.
In line with our experimental findings, we note that a unit amount of Joule heat, generated at increasingly large separations from the temperature probe, leads to ever smaller $\Delta T$-signals, resulting thereby into ever higher $K$-values. Clearly, employing the $K^{\mathrm{c.e.}}$--- calibrated with Joule heat generated at the electrical contacts--- in a calculation determining the reversible heat generated at a location closer to temperature probe, will overestimate its value. For the main text, we therefore chose to prioritize  $K^{\mathrm{h.e.}}$.

The above calibration uncertainty would not have been present if the cell had had a uniform internal temperature at each instance. 
To accomplish that, instead of a sudden step, one should \textit{slowly} apply the external potential, such that the cell can equilibrate internally while being heated. Such experiments with the current cell were unsuccessful because heat flowed out of the cell too quickly; the measured temperature differences were too small compared to the equilibrium temperature fluctuations to separate signal from noise. 
Hence, resolving the calibration uncertainty poses an experimental challenge, which will require the development of a better insulated and an even more sensitive calorimetric setup. 

All above-mentioned reservations aside, the obtained heat transfer coefficients $K^{\mathrm{c.e.}}$ and $K^{\mathrm{h.e.}}$ still only differ by approximately $30\%$.

\section{Time scales of current and temperature-difference decay}\label{sec:app_timescales}
Figure~\ref{fig3}(b) and (c) exhibit a decaying current and temperature difference that we now discuss in more detail.

In general, electric double layer capacitors (EDLCs) are known to display a richer charging dynamics than conventional RC-circuits because  both the capacitance and electrical resistance of an EDLC depend on its charging state. For instance, model calculations on a parallel plate EDLC (plate separation $L$) showed that 
surface charge is screened on a time scale $\ld L/D$ (with $D$ the diffusion constant), but corrections also introduce other time scales: $\lambda_{S}L/D$, $\lambda_{S}\ld/D$, $\ld^{2}/D$ (with $\lambda_{S}$ the thickness of a Stern layer) \cite{bazant2004diffuse}. At later times ($L^2/D$) salt diffusion takes over.

\begin{figure}
\def\svgwidth{0.49\textwidth}{
 \providecommand\color[2][]{%
    \errmessage{(Inkscape) Color is used for the text in Inkscape, but the package 'color.sty' is not loaded}%
    \renewcommand\color[2][]{}%
  }%
  \providecommand\transparent[1]{%
    \errmessage{(Inkscape) Transparency is used (non-zero) for the text in Inkscape, but the package 'transparent.sty' is not loaded}%
    \renewcommand\transparent[1]{}%
  }%
  \providecommand\rotatebox[2]{#2}%
  \ifx\svgwidth\undefined%
    \setlength{\unitlength}{611.42745019bp}%
    \ifx\svgscale\undefined%
      \relax%
    \else%
      \setlength{\unitlength}{\unitlength * \real{\svgscale}}%
    \fi%
  \else%
    \setlength{\unitlength}{\svgwidth}%
  \fi%
  \global\let\svgwidth\undefined%
  \global\let\svgscale\undefined%
  \makeatother%
  \begin{picture}(1,0.48)%
    \put(0,0){\includegraphics[width=\unitlength]{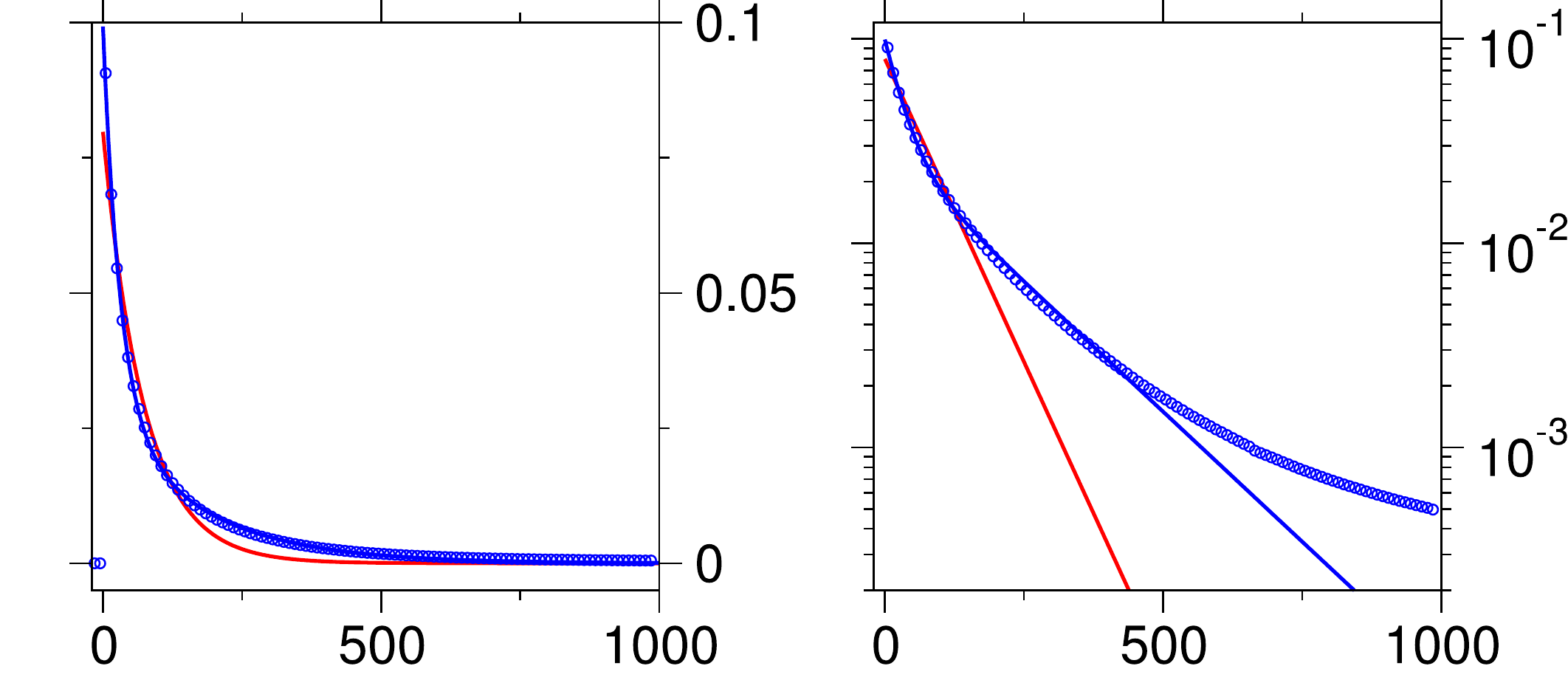}}%
    \put(0.01,0.23101035){\color[rgb]{0,0,0}\makebox(0,0)[lb]{\smash{\begin{rotate}{90} $I$ [A]\end{rotate}}}}%
    \put(0.23253116,-0.02){\color[rgb]{0,0,0}\makebox(0,0)[lb]{\smash{t [s]}}}%
    \put(0.73177572,-0.02){\color[rgb]{0,0,0}\makebox(0,0)[lb]{\smash{t [s]}}}%
          \put(0.0,0.45){\color[rgb]{0,0,0}\makebox(0,0)[lb]{\smash{(a)}}}%
    \put(0.5,0.45){\color[rgb]{0,0,0}\makebox(0,0)[lb]{\smash{(b)}}}%
  \end{picture}%
}
\caption{\label{fig_sup5}%
 The current relaxation data of the left panel of Fig.~\ref{fig3}(b)  replotted on linear (a) and log-linear scale (b) with blue circles (only one tenth of the data is shown). Fits $I_{\#1}(t)$ (red line) and $I_{\#2}(t)$ (blue line) as introduced in the text are shown as well.}
\end{figure}
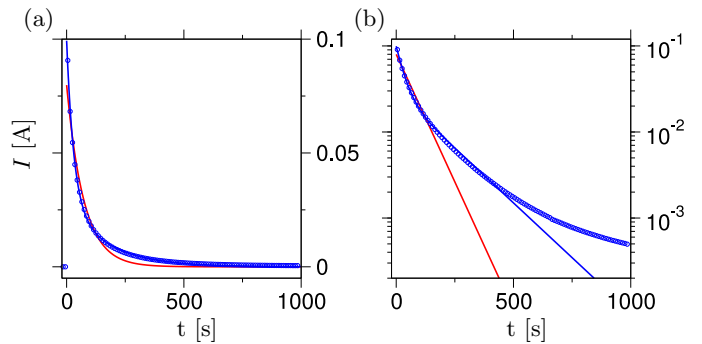

The electrodes of this Letter, made from porous carbon, have a hierarchical structure in which  more length scales must come into play. One can therefore expect to find an even richer charging dynamics. As such, it comes as no surprise that data of the left panel of Fig.~\ref{fig3}(b) --- redrawn in Fig.~\ref{fig_sup5} with blue circles on linear (a) and log-linear (b) scale--- 
 confirm the picture sketched above of the absence of a single time scale to the decay of the current; clearly, the data is poorly fitted by a single decaying exponential (red line) $I_{\#1}(t)=I_{1}\exp[-t/\tau_{1}]$ with $I_{1}=0.080$ A and $\tau_{1}=73$ s. Adding a second exponential $I_{\#2}(t)=I_{1}\exp[-t/\tau_{1}]+I_{2}\exp[-t/\tau_{2}]$ with $I_{0}=0.071$ A, $I_{1}=0.028$ A, $\tau_{0}=31$~s, and $\tau_{0}=171$~s) fits the data better (blue line) at short times but still fails at long times. 

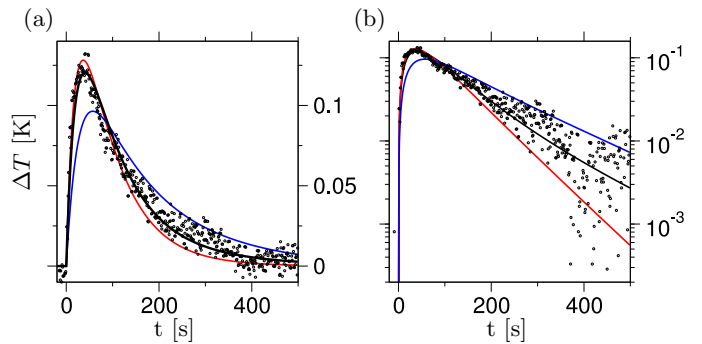
\begin{figure}
\def\svgwidth{0.49\textwidth}{
  \providecommand\color[2][]{%
    \errmessage{(Inkscape) Color is used for the text in Inkscape, but the package 'color.sty' is not loaded}%
    \renewcommand\color[2][]{}%
  }%
  \providecommand\transparent[1]{%
    \errmessage{(Inkscape) Transparency is used (non-zero) for the text in Inkscape, but the package 'transparent.sty' is not loaded}%
    \renewcommand\transparent[1]{}%
  }%
  \providecommand\rotatebox[2]{#2}%
  \ifx\svgwidth\undefined%
    \setlength{\unitlength}{606.60228725bp}%
    \ifx\svgscale\undefined%
      \relax%
    \else%
      \setlength{\unitlength}{\unitlength * \real{\svgscale}}%
    \fi%
  \else%
    \setlength{\unitlength}{\svgwidth}%
  \fi%
  \global\let\svgwidth\undefined%
  \global\let\svgscale\undefined%
  \makeatother%
  \begin{picture}(1,0.48)%
    \put(0,0){\includegraphics[width=\unitlength]{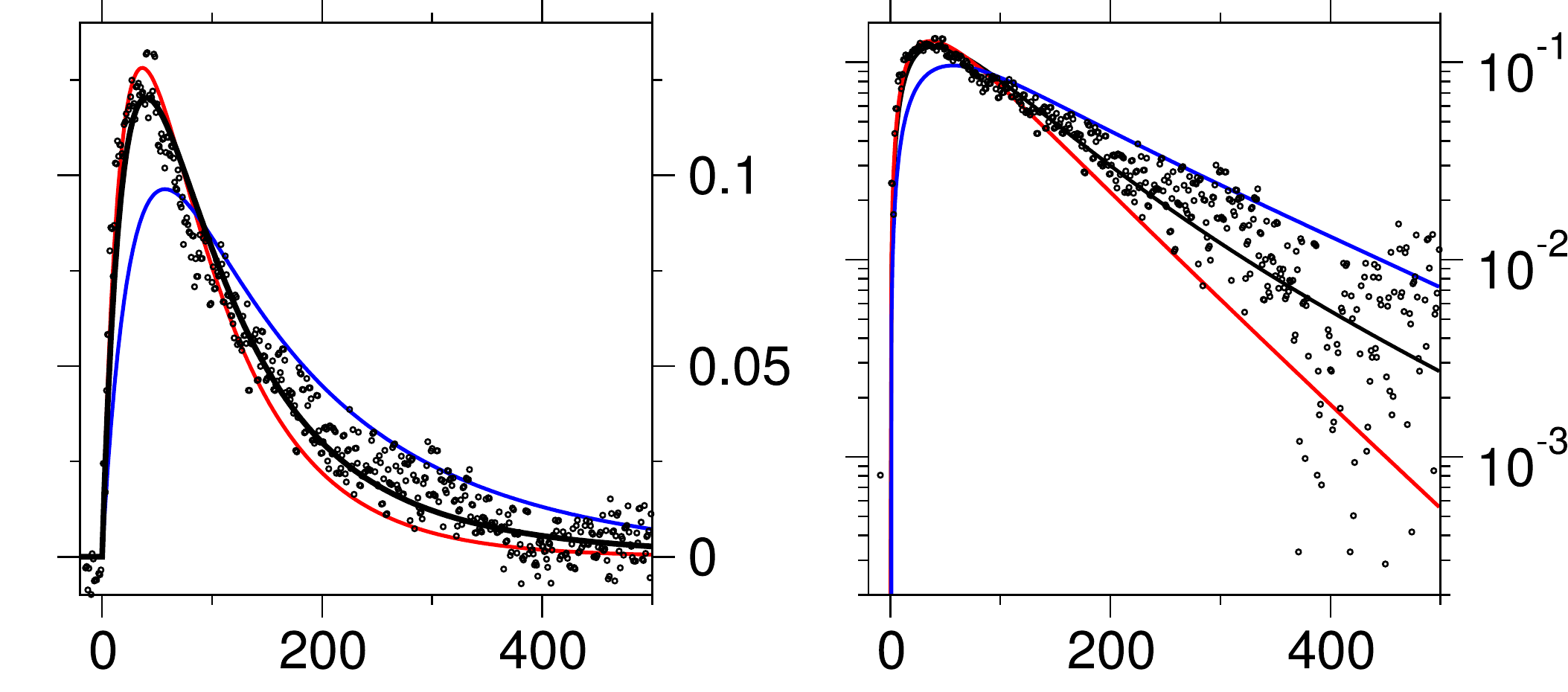}}%
    \put(0.01,0.2){\color[rgb]{0,0,0}\makebox(0,0)[lb]{\smash{\begin{rotate}{90} $\Delta T$ [K]\end{rotate}}}}%
    \put(0.19785768,-0.02){\color[rgb]{0,0,0}\makebox(0,0)[lb]{\smash{t [s]}}}%
    \put(0.70107342,-0.02){\color[rgb]{0,0,0}\makebox(0,0)[lb]{\smash{t [s]}}}%
              \put(0.0,0.45){\color[rgb]{0,0,0}\makebox(0,0)[lb]{\smash{(a)}}}%
    \put(0.5,0.45){\color[rgb]{0,0,0}\makebox(0,0)[lb]{\smash{(b)}}}%
  \end{picture}%
}
\caption{\label{fig_sup6}%
 The thermal response data of the left panel of Fig.~\ref{fig3}(c) replotted on linear (a) and log-linear scale (b) with black circles.
 Fits to these data are obtained with Eq.~(\ref{eq.sol}) for different assumptions about $\dot{\Pi}_{\rm tot}$: Joule heat only (red line), reversible heat only (blue line), and both Joule and reversible heat (black line).}
\end{figure}

In Fig.~\ref{fig_sup6} we reproduce (with black circles) the data of the left panel of Fig.~\ref{fig3}(c) on linear (a) and log-linear (b) scale. To model these data, we note that Eq.~\eqref{eq:heatequation} allows for the following general solution
\begin{equation}\label{eq.sol}
\Delta T(t)=\exp[-t/\tau]\left[\int^{t}_{0}\exp[t'/\tau]\frac{\dot{\Pi}_{\rm tot}(t')}{\mathbb{C}_{p}}dt' \right],
\end{equation}
with $\dot{\Pi}_{\rm tot}(t)$ the heating rate and $\tau=\mathbb{C}_{p}/K=61.1\pm0.3$~s the thermal time constant as determined in the previous section. Clearly, the time scales of thermal response to charging are dictated both by $\tau$ and by the time scales with which $\dot{\Pi}_{\rm tot}$ decays. In the main text we argued that $\dot{\Pi}_{\rm tot}$ contains Joule heat contributions ($\sim I^2$) and reversible contributions ($\sim I$):
we consider  the following general form  $\dot{\Pi}_{\rm tot}(t)=\left[\zeta_{\rm irr}I(t)^2+\zeta_{\rm rev}I(t)\right]$ where $\zeta_{\rm irr}$ and  $\zeta_{\rm rev}$ are parameters of dimension $\Omega$ and $V$, respectively.
From the above discussion we know that the current decay is characterized by multiple time scales. However, to proceed, we approximate $I(t)$ with the fit found above: $I(t)=\Theta(t) I_{\#2}(t)$, with $\Theta(t) $ a Heaviside step function. Equation (\ref{eq.sol}) can now be solved analytically: a lengthy expression with two free parameters $\zeta_{\rm irr}$ and $\zeta_{\rm rev}$.

To start, we consider the case of only Joule heat by setting $\zeta_{\rm rev}=0$ V. The best fit (red line) is then found for $\zeta_{\rm irr}=1.1$~$\Omega$. Conversely, considering the case of only reversible heat (setting $\zeta_{\rm irr}=0$~$\Omega$) leads to a best fit (blue) at $\zeta_{\rm rev}=0.051$~V. The short-time $\Delta T$ response is decently described by the case of only Joule heat, while the case of only reversible heat  performs poorly in that region.
Figure~\ref{fig_sup6}(b) shows that both choices do not reproduce the long-time relaxation accurately: they relax either too rapidly (only Joule heat) or too slowly (only reversible heat).


As can be expected on the basis of Ref.~\cite{janssen2017reversible} and this Letter, the physical situation is that of reversible and irreversible heat {\it simultaneous} at play.
Indeed, the best fit through the data in Fig.~\ref{fig_sup6} is achieved for $\zeta_{\rm irr}=0.76$~$\Omega$ and  $\zeta_{\rm rev}=0.016$ V; hence, a mixture of reversible and reversible contributions. We conclude that, in order to reproduce the long-time relaxation correctly, it seems necessary to account for both reversible and irreversible heat. 

So far, we only considered $\zeta_{\rm rev/irr}$ to be time-independent. However, for the slit geometry of Ref.~\cite{janssen2017reversible}, $\zeta_{\rm rev}$ takes the form $\zeta_{\rm rev}=\kbt (\partial_{z}q)/[e (\rho_{+}+\rho_{-})]$; hence, there is no reason to believe $\zeta_{\rm rev}$ is truly time independent. Relaxing this assumption could lead to an even better reproduction of the measured temperature data.

\end{appendix}

\end{document}